\documentclass[12pt,twoside]{article}
\usepackage[makeroom]{cancel}
\usepackage{latexsym}
\usepackage{longtable}
\usepackage{epsfig}
\usepackage{graphicx,bbm,psfrag}
\usepackage{amssymb,amsmath,amsthm,booktabs,mathtools}
\usepackage[subnum]{cases}
\usepackage{bm}
\usepackage{color}

\usepackage[english]{babel}
\usepackage[utf8x]{inputenc}
\usepackage[T1]{fontenc}

\newcommand{\vx}{\vec{x}}
\newcommand{\vy}{\vec{y}}
\newcommand{\vp}{\vec{p}}
\newcommand{\vq}{\vec{q}}
\newcommand{\Ep}{E_{\vec{p}}}
\newcommand{\Epp}{E_{\vec{p}^{\prime}}}
\newcommand{\Eq}{E_{\vec{q}}}
\newcommand{\Er}{E_{\vec{r}}}
\newcommand{\Op}{\omega_{\vec{p}}}
\newcommand{\aps}{a_{\vec{p}}^s}

\newcommand{\aqr}{a_{\vec{q}}^r}
\newcommand{\bps}{b_{\vec{p}}^s}

\newcommand{\bqr}{b_{\vec{q}}^r}
\newcommand{\adps}{a_{\vec{p}}^{s\dagger}}

\newcommand{\adqr}{a_{\vec{q}}^{r\dagger}}
\newcommand{\bdps}{b_{\vec{p}}^{s\dagger}}

\newcommand{\bdqr}{b_{\vec{q}}^{r\dagger}}
\newcommand{\Aps}{A_{\vec{p}}^s}

\newcommand{\Aqr}{A_{\vec{q}}^r}
\newcommand{\Bps}{B_{\vec{p}}^s}

\newcommand{\Bqr}{B_{\vec{q}}^r}
\newcommand{\Adps}{A_{\vec{p}}^{s\dagger}}

\newcommand{\Adqr}{A_{\vec{q}}^{r\dagger}}
\newcommand{\Bdps}{B_{\vec{p}}^{s\dagger}}

\newcommand{\Bdqr}{B_{\vec{q}}^{r\dagger}}

\newcommand{\Dp}{\frac{d^3p}{(2\pi)^3}}

\newcommand{\deltasss}{\hat{\delta}^{(3)}(\vp-\vq)}
\newcommand{\deltas}{\hat{\delta}_t^{(3)}(\vp-\vq)}
\newcommand{\deltass}{\hat{\delta}_t^{\prime (3)}(\vp-\vq)}
\newcommand{\Us}{\bar{u}^{s^{\prime}}(p)\gamma^1u^s(q)}
\newcommand{\Vs}{\bar{v}^{s^{\prime}}(q)\gamma^1v^s(p)}
\newcommand{\betal}{\beta_{\mathrm{L}}}
\newcommand{\betar}{\beta_{\mathrm{R}}}
\newcommand{\mul}{\mu_{\mathrm{L}}}
\newcommand{\mur}{\mu_{\mathrm{R}}}
\newcommand{\me}{m_{\mathrm{eff}}}

\newcommand{\ii}{\mathrm{i}}

\newcommand{\n}{\nonumber \\}

\makeatletter

\begin{document}
\begin{center}
{\Large {\bf Full counting statistics in the free Dirac theory}}

\vspace{1cm}

{\large Takato Yoshimura$^{1,2}$}
\vspace{0.2cm}

{\small\em
$^1$ Department of Mathematics, King's College London, Strand, London WC2R 2LS, U.K.}\\
{\small\em
$^2$  Institut de Physique Th\`eorique, CEA Saclay, Gif-Sur-Yvette, 91191, France}
\end{center}

\vspace{1cm}

\noindent We study charge transport and fluctuations of the (3+1)-dimensional massive free Dirac theory. In particular, we focus on the steady state that emerges
 following a local quench whereby two independently thermalized halves
 of the system are connected and let to evolve unitarily for a long
 time. Based on the two-time von Neumann measurement statistics and exact computations, the scaled cumulant generating function 
 associated with the charge transport is derived. We find that it can be written as a generalization of Levitov-Lesovik formula to the case in
 three spatial dimensions. In the massless case, we note that only the first four scaled cumulants are nonzero. Our results provide also a direct confirmation for the validity of the
 extended fluctuation relation in higher
 dimensions. An application of our approach to Lifshitz fermions is also briefly discussed.

\newpage

\tableofcontents

\newpage

\section{Introduction}
Our understanding on transport phenomena in quantum many-body systems has been significantly advanced by fruitful interactions between theory and experiment over the past decades. There are many possible situations to study transport phenomena, and one of the simplest protocols would be a local quench where a non-equilibrium steady state (NESS) is generated upon gluing two systems, which are initially prepared at different parameters (say chemical potentials or temperatures). This so-called {\it partitioning protocol} has gained a surge of interest in recent years \cite{BDreview,VMReview} leading to studies on a variety of models in the protocol, from one-dimensional integrable systems to higher dimensional quantum critical systems \cite{BD2012,DeLucaVitiXXZ,Nat-Phys,CM,doyonKG,rarefact1,rarefact2,CDY,bertini1,prosen1}. That being said, it is important to note that these studies have focused mainly on one-dimensional systems due to the abundance of available analytical approaches and simulability by powerful numerical methods such as tDMRG in one dimension. This is in stark contrast with the case in higher dimensions where relatively less is known. This case also applies to the study of charge fluctuation (a.k.a. full counting statistics) in which a significant amount of works have been done on one-dimensional electron (impurity) systems \cite{klich,BN,GK,schonhammer,GGM,BD1,BD2}, initiated by a seminal work by Levitov and Lesovik \cite{LL1,LL2} in 90s. In view of these situations, it is urgent to reinforce our understanding on the non-equilibrium dynamics and fluctuations in higher dimensions. In this work, we present a first detailed analysis on the full counting statistics in the (3+1)-D Dirac theory using the partitioning protocol. On the experimental side, for the last decade, a surprising ubiquity of Dirac fermions in nature, such as graphene \cite{graphene} and Dirac semimetal \cite{dirac_semimetal} has been realized. It is therefore paramount to understand its transport nature in one of the simplest and cleanest situations which could serve as a benchmark.

It is worth recalling what has been done concerning the studies on non-equilibrium transport in higher dimensions. Being initiated by the work \cite{Nat-Phys}, studies on NESS in higher dimensions have been addressed by AdS/CFT correspondence (for quantum critical systems) \cite{Nat-Phys}, exact computations (for free models) \cite{CM,doyonKG}, and hydrodynamics \cite{rarefact1,rarefact2}, which, in general, plays a pivotal role in studying transport phenomena. In particular, while the quantities of primary interest are the space-time profiles of the local density and current in NESS, the fluctuation in energy transport and the approach of observables towards NESS in the higher-D Klein-Gordon model were also studied in \cite{doyonKG} by making use of powerful free field techniques. The techniques developed there will be intensively exploited in this paper as well.

The present paper deals with yet another representative free model, the (3+1)-D free Dirac model, and discusses the charge fluctuation in the NESS generated by a local quench. Since the model possesses two charged particles, Dirac fermions and anti-Dirac fermions, we expect that there occurs a $U(1)$-charge flow when two systems with different global parameters are put in contact. A local quench we consider can be regarded as a simpler version of the protocol used in the study of full counting statistics: the total Hamiltonian that does time-evolution now is without impurities, thus particles are transferred without reflection at the junction (i.e. the transmission coefficient is simply 1). A typical quantity that encodes all the information about the charge fluctuation is the {\it scaled cumulant generating function} (SCGF), which is a generating function for the probability distribution of the transferred $U(1)$-charge across the contact surface for the duration of large time $t\to\infty$ (the infinite system size limit $L\to\infty$ is taken first so that the propagation front does not reach the boundary). There are two situations for which we can compute the SCGF. One is to prepare the initial state $\rho_{\rm L}\otimes\rho_{\rm R}$, where $\rho_{\rm L}$ and $\rho_{\rm R}$ are the density operators of the left and right baths, at $t=0$ and focus on the charges transferred between $t=0$ and $t\to\infty$. In this situation, the counted charges are not necessarily carried by the NESS as the NESS is reached only at large time. Another case is where the initial state is prepared at $t=t_0\to-\infty$ so that at $t=0$ the NESS emerges across the interface. The system is then fully in the NESS between time $t=0$ and $t\to\infty$ during which the transferred charges are counted, and this is the scenario we shall be interested in. In order to assess the SCGF, we first derive the stationary density operator that describes NESS. This allows us to explicitly evaluate the SCGF based on the two-time von Neumann measurement statistics where computations are carried out by free field techniques \cite{doyonKG}. It is found that the so-obtained SCGF is nothing but a straightforward generalization of the Levitov-Lesovik formula for (3+1)-D Dirac theory, implying the universality of the formula regardless of the dimensionality. Interestingly, the SCGF  has a particularly simple form in the massless limit, and it turns out that the cumulants exist only up to fourth order. We also check the validity of the {\it extended fluctuation theorem} (EFR) proposed in \cite{BD3} for the SCGF, which is expected to hold in free theories and 2D conformal field theories.
  

\section{Non-equilibrium steady-state of the model}
\subsection{The Dirac model}
In order to introduce our notation, we briefly recall the canonical
quantization of the Dirac model. The Hamiltonian of the relativistic Dirac theory is given by
\begin{equation}\label{hamil}
 H = \int d^3x :\psi^{\dagger}(-\ii\gamma^0 \vec{\gamma} \cdot \vec{\nabla} + m\gamma^0)\psi:,
\end{equation}
where the fermionic field operators are defined as
\begin{align}\label{a-rep}
 \psi(\vec{x},t) &=\int Dp\sum_s
 (\aps u^s(p)e^{-\ii p\cdot x}+\bdps v^s(p)e^{\ii p\cdot x}) \\
 \overline{\psi}(\vec{x},t) &= \int Dp\sum_s
 (\bps\bar{v}^s(p)e^{-\ii p\cdot x}+\adps\bar{u}^s(p)e^{\ii p\cdot x}),
\end{align}
with $p\cdot x=p^{\mu}x_{\mu}$ ($g_{\mu\nu}=\mathrm{diag}(1,-1,-1,-1)$) and $Dp=d^3p/(2\pi)^3\sqrt{2\Ep}$. Furthermore, spinors $u^s(p)$ and
$v^s(p)$ are expressed as
\begin{equation}
 u^s(p)=\left(\begin{array}{c}
  \sqrt{p\cdot \sigma}\xi^s \\
	 \sqrt{p\cdot \bar{\sigma}}\xi^s
	\end{array}\right),
\ \ v^s(p)=\left(\begin{array}{c}
  \sqrt{p\cdot \sigma}\eta^s \\
	 -\sqrt{p\cdot \bar{\sigma}}\eta^s
	\end{array}\right),
\end{equation}
where $\xi^s$ and $\eta^s$ are two different bases of two-component
spinors, and $\sigma^{\mu}=(\hat{1},\vec{\sigma})$ and
$\bar{\sigma}^{\mu}=(\hat{1},-\vec{\sigma})$. The creation and annihilation operators satisfy the anti-commutation
relation rules
\begin{equation}\label{commu}
\{\aps,\adqr \} = \{ \bps, \bdqr \} = \deltasss \delta^{rs}
 \end{equation}
 where we defined $\deltasss=(2\pi)^3\delta^{(3)}(\vp-\vq)$, and with all other anti-commutators being zero. This implies equal-time anti-commutation relations
\begin{align}
 \{\psi_a(\vx),\psi_b^{\dagger}(\vy)\} = \delta^{(3)}(\vx-\vy)\delta_{ab}, \{\psi_a(\vx),\psi_b(\vy)\} =  \{\psi_a^{\dagger}(\vx),\psi_b^{\dagger}(\vy)\} = 0.
\end{align}
The initial density matrix is simply defined as a (tensor) product of
sub-density matrices representing subsystems in thermal equilibrium with
different temperatures and chemical potentials:
\begin{equation}
 \rho_{\mathrm{th}} = \exp[-\beta_{\mathrm{L}}(H_{\mathrm{L}}-\mu_{\mathrm{L}}Q_{\mathrm{L}})-\beta_{\mathrm{R}}(H_{\mathrm{R}}-\mu_{\mathrm{R}}Q_{\mathrm{R}})],
\end{equation}
where $H_{\mathrm{L},\mathrm{R}}$ and $Q_{\mathrm{L},\mathrm{R}}$ are
Hamiltonians and $U(1)$ charges for subsystems
defined as
\begin{align}
H_{\mathrm{L},\mathrm{R}} &= \int_{x_1\lessgtr 0} d^3x
 :\psi^{\dagger}(-\ii\gamma^0 \vec{\gamma}
 \cdot \vec{\nabla} + m\gamma^0)\psi: \label{semihamil}\\ 
 Q_{\mathrm{L},\mathrm{R}} &= \int_{x_1\lessgtr 0}
  d^3x\psi^{\dagger}(\vx)\psi(\vx),
\end{align}
and with $\mu_{\mathrm{L},\mathrm{R}}$ being chemical potentials for
semi-halves. The total $U(1)$ charge can be expressed in terms of creation and
annihilation operators
\begin{equation}
 Q=\int \Dp \sum_s (\adps \aps - \bdps \bps).
\end{equation}
\subsection{NESS density operator}
In order to analyze the charge fluctuation (large deviation) of the Dirac theory, we need an explicit expression of the NESS density
operator $\rho_{\rm s}$, which is defined by, for a generic field $\mathcal{O}$,
\begin{equation}
\mathrm{Tr}(\mathfrak{n}[\rho_{\rm s}]\mathcal{O})=\lim_{t \to \infty}\mathrm{Tr}(\mathfrak{n}[\rho_{\rm th}]e^{\ii Ht}\mathcal{O}e^{-\ii Ht}),
\end{equation}
where $\mathfrak{n}[\rho]=\rho/\mathrm{Tr}\rho$ for $\rho=\rho_{\rm s}, \rho_{\rm th}$.
In what follows, we will show that the explicit form of $\rho_{\rm s}$ reads
\begin{align}
 \rho_{\mathrm{s}} &=
 \exp[-\beta_{\mathrm{L}}(H_{\mathrm{L}}^+-\mu_{\mathrm{L}}Q_{\mathrm{L}}^+)-\beta_{\mathrm{R}}(H_{\mathrm{R}}^+-\mu_{\mathrm{R}}Q_{\mathrm{R}}^+)]
 \nonumber \\
 &= \exp \Bigl[-\int\frac{d^3p}{(2\pi)^3}\sum_s(V_+(\vp)\adps
  \aps + V_-(\vp)\bdps \bps)\Bigr], \label{ness}
\end{align}
with
\begin{align}
 H_{\mathrm{L},\mathrm{R}}^+&=\int_{p^1\gtrless 0}\Dp E_{{\bf p}}\sum_s (\adps \aps + \bdps \bps) \\
 Q_{\mathrm{L},\mathrm{R}}^+&=\int_{p^1\gtrless 0}\Dp \sum_s (\adps \aps - \bdps \bps),
\end{align}
and where
\begin{equation}
 V_+(\vp) = \begin{cases}
		 \beta_{\mathrm{L}}(E_{\vp}-\mu_{\mathrm{L}}) & \text{$p^1>0$} \\
		 \beta_{\mathrm{R}}(E_{\vp}-\mu_{\mathrm{R}}) & \text{$p^1<0$}
		\end{cases},\ \ 
 V_-(\vp) = \begin{cases}
		 \beta_{\mathrm{L}}(E_{\vp}+\mu_{\mathrm{L}})& \text{$p^1>0$} \\
		 \beta_{\mathrm{R}}(E_{\vp}+\mu_{\mathrm{R}})& \text{$p^1<0$}
		\end{cases}.
\end{equation}
For later convenience, we note that the NESS density operator $\rho_{\rm s}$ satisfies the following contraction relations
\begin{align}\label{contract1}
 \begin{split}
 \mathrm{Tr}(\mathfrak{n}[\rho_{\mathrm{s}}]\adps \aqr) &=
 \frac{\deltasss\delta^{rs}}{1+e^{V_+(\vec{p})}},\quad\mathrm{Tr}(\mathfrak{n}[\rho_{\mathrm{s}}]\bdps \bqr) =
 \frac{\deltasss\delta^{rs}}{1+e^{V_-(\vec{p})}} \\
\mathrm{Tr}(\mathfrak{n}[\rho_{\mathrm{s}}]\aps \adqr) &=
 \frac{\deltasss\delta^{rs}}{1+e^{-V_+(\vec{p})}},\quad\mathrm{Tr}(\mathfrak{n}[\rho_{\mathrm{s}}]\bps \bdqr) =
 \frac{\deltasss\delta^{rs}}{1+e^{-V_-(\vec{p})}}.
 \end{split}
\end{align}

The following derivation will closely parallel the argument employed in \cite{doyonKG}. First let us introduce the ``B-representation'' of the fundamental fields. Recall that the representation \eqref{a-rep} with \eqref{commu}, which we call the ``A-representation'' as in \cite{doyonKG}, diagonalizes the total Hamiltonian \eqref{hamil}. The B-representation instead diagonalizes Hamiltonians of each half \eqref{semihamil}. With a natural choice of a boundary condition $\psi_{\rm B}(x^1=0,\tilde{x})=0$, where $\tilde{x}=(x^2,x^3)$, the Dirac fields in the B-representation read  
\begin{align}\label{b-rep}
 \psi_{\rm B}(\vx) &=\int Dp\sum_s
 (\Aps u^s(p)e^{-i\vec{p}\cdot \vec{x}}+\Bdps v^s(p)e^{ip\cdot x})(\vartheta(p^1x^1)-\vartheta(-p^1x^1)) \\
 \overline{\psi}_{\rm B}(\vx) &= \int Dp\sum_s
 (\Bps\bar{v}^s(p)e^{-ip\cdot x}+\Adps\bar{u}^s(p)e^{ip\cdot x})(\vartheta(p^1x^1)-\vartheta(-p^1x^1)),
\end{align}
where $\Aps, \Adps, \Bps$, and $\Bdps$ satisfy the same anti-commutation relation \eqref{commu} and
\begin{align}
 H_{\mathrm{L},\mathrm{R}}&=\int_{p^1\gtrless 0}\Dp E_{{\bf p}}\sum_s (\Adps \Aps + \Bdps \Bps) \\
 Q_{\mathrm{L},\mathrm{R}}&=\int_{p^1\gtrless 0}\Dp \sum_s (\Adps \Aps - \Bdps \Bps).
\end{align}
This implies the following contraction rules 
\begin{align}\label{contract2}
\begin{split}
 \mathrm{Tr}(\mathfrak{n}[\rho_{\rm th}]\Adps \Aqr) &=
 \frac{\deltasss\delta^{rs}}{1+e^{V_+(\vec{p})}},\quad \mathrm{Tr}(\mathfrak{n}[\rho_{\rm th}]\Bdps \Bqr) =
 \frac{\deltasss\delta^{rs}}{1+e^{V_-(\vec{p})}} \\
\mathrm{Tr}(\mathfrak{n}[\rho_{\rm th}]\Aps \Adqr) &=
 \frac{\deltasss\delta^{rs}}{1+e^{-V_+(\vec{p})}},\quad
 \mathrm{Tr}(\mathfrak{n}[\rho_{\rm th}]\Bps \Bdqr) =
 \frac{\deltasss\delta^{rs}}{1+e^{-V_-(\vec{p})}},
\end{split}
\end{align}
which are nothing but the rules \eqref{contract1} replacing $\rho_{\rm s}$, $\aps$ (resp. $\bps$) and $\adps$ (resp. $\bdps$) with $\rho_{\rm s}$, $\Aps$ (resp. $\Bps$) and $\Adps$ (resp. $\Bdps$). In fact, we will observe that, for any operator $\mathcal{O}_{\rm A}$ in the A-representation,
\begin{equation}\label{sop}
\lim_{t \to \infty}\mathrm{Tr}(\mathfrak{n}[\rho_{\rm th}]e^{\ii Ht}\mathcal{O}_{\rm A}e^{-\ii Ht})=\mathrm{Tr}(\mathfrak{n}[\rho_{\rm th}]S(\mathcal{O}_{\rm A})).
\end{equation}
Here $S$ is the scattering isomorphism operator that satisfies
\begin{equation}
S(\aps)=\Aps,\quad S(\adps)=\Adps,\quad S\Bigl(\prod_{p,s}(\aps)^{\eta_{p,s}}\Bigr)=\prod_{p,s}S((\aps)^{\eta_{p,s}}),
\end{equation}
where analogous relations hold for $\bps$ and $\Bps$ (and their hermite conjugates) as well. To demonstrate \eqref{sop}, it is sufficient to show that
\begin{equation}
\psi_{\rm B}(\vx,t)=\lim_{t\to\infty}S(\psi_{\rm A}(\vx,t)) + \Psi(\vx,t),
\end{equation}
where the correction $\Psi(\vx,t)$ has no contribution in evaluating averages of any physical operator at $t\to\infty$. A proof of this equality is provided in Appendix \ref{bconv}.

On a physical basis, essentially, the relation between
$\rho_{\mathrm{s}}$ and $\rho_{\mathrm{th}}$ can be attributed to the existence of the M\o ller operator \cite{taylor} $S_+=\lim_{t\to \infty}e^{-itH}e^{itH_0}$ that intertwines $|\phi\rangle_0$ and $|\phi\rangle$ as
$|\phi\rangle=S_+|\phi\rangle_0$, where $|\phi\rangle_0$ and $|\phi\rangle$ are eigenstates of Hamiltonians
$H_0=H_{\mathrm{L}}+H_{\mathrm{R}}$ and $H$ respectively\footnote{This relation is valid upon being
evaluated in matrix elements.}. Assuming that the spectra of those
Hamiltonians are same, we can obtain $\rho_{\mathrm{s}}$ from
$\rho_{\mathrm{th}}$. More intuitively, we can argue using the wave packet language: wave packets carrying
positive momenta must, in the far past, come from the left half so that
they can have information only about left side. The same argument holds for wave packets going towards left side. Notice that this reasoning is
true only when the traveling wave packets are not affecting each other
in the course of time-evolution
\cite{Doyon1}.

\section{Charge fluctuations}
\subsection{Measurement statistics}
Of primary interest in this paper is the statistics of
transferred charges $Q_{\rm tra}=\frac{1}{2}(Q_R-Q_L)$ in the NESS reached
after a long time. We shall consider a von
Neumann measurement of this quantity in the NESS regime: let the system be prepared at
$t=t_0<0$ with the density matrix $\rho_{\mathrm{th}}=\rho_{\rm L}\otimes\rho_{\rm R}$ with the initial Hilbert space $\mathcal{H}_0=\mathcal{H}_{\rm L}\otimes\mathcal{H}_{\rm R}$, and evolve
unitarily with the full Hamiltonian $H$ after $t_0$. Suppose then we measure $q_0$ at $t=0$
and $q_t$ at $t$: the system is assumed to support the NESS in between the light-cone
even when $t=0$ as we take
a limit $t_0\to -\infty$ later. A joint probability of those measurements is
\begin{equation}
 P(q_t;q_0) = \mathrm{Tr}\Bigl(P_{q_t}U_tP_{q_0}\rho_0P_{q_0}U^{\dagger}_tP_{q_t}\Bigr),
\end{equation}
where we defined the density matrix at $t=0$ as
$\rho_0=U_{t_0}^{\dagger}\rho_{\mathrm{th}}U_{t_0}$ with $U_t=e^{-\ii Ht}$,
and $P_a=|a\rangle \langle
a|$ are projection operators. Furthermore, the generating function for the
charge transfer associated with this probability is
defined as
\begin{equation}
 P_{t_0}(\lambda;t)=\sum_{q_t,q_0=-\infty}^{\infty}e^{\ii\lambda(q_t-q_0)}P(q_t;q_0).
\end{equation}
Noting that the eigenvalues of $Q_{\rm tra}$ are either half integers or integers, one can rewrite
the generating function with introducing a dummy variable $\gamma$ \cite{EHM}:
\begin{align}
 P_{t_0}(\lambda;t)&=\int_0^{4\pi}\frac{d\gamma}{4\pi}\mathrm{Tr}_{\mathcal{H}_0}\Bigl(\rho_0e^{\ii(\gamma-\lambda/2)Q_{\rm tra}}e^{\ii\lambda
  Q_{\rm tra}(t)}e^{-\ii(\gamma+\lambda/2)Q_{\rm tra}}\Bigr) \nonumber \\
 &=\int \frac{d\gamma}{4\pi}\mathrm{Tr}_{\mathcal{H}_0}(\rho_0e^{\Theta_{\gamma,\lambda}(t)}),
\end{align}
where $Q_{\rm tra}(t)=e^{\ii Ht}Q_{\rm tra}e^{-\ii Ht}$, and
\begin{equation}
 e^{\Theta_{\lambda,\gamma}(t)}:=e^{\ii(\gamma-\lambda/2)Q_{\rm tra}}e^{\ii\lambda
  Q_{\rm tra}(t)}e^{-\ii(\gamma+\lambda/2)Q_{\rm tra}}.
\end{equation}
The generating function in
a stationary regime, which is obtained by the limit $t_0
\to -\infty$, is then
\begin{equation}
 P_{\mathrm{s}}(\lambda;t) = \lim_{t_0\to
  -\infty}P_{t_0}(\lambda;t)= \int
  \frac{d\gamma}{4\pi}\mathrm{Tr}(\rho_{\mathrm{s}}e^{\Theta_{\lambda,\gamma}(t)}). \label{eq:gf}
\end{equation}
In fact, we have to `scale' it in order to have a finite
charge transfer as we are dealing with the infinite system. Hence
assuming transverse directions (i.e. $x^2$ and $x^3$) have a linear size $L$,
we shall evaluate the following instead of the above:
\begin{equation}
 F(\lambda)=\lim_{t,L\to
  \infty}\frac{1}{tL^2}P_{\mathrm{s}}(\lambda;t), \label{eq:scgf}
\end{equation}
where the limit is taken with $t\ll L$ (the speed of light is set to $c=1$). 
Note further that, since $Q_{\rm tra}$ and
$Q_{\rm tra}(t)$ are bilinears in fermion operators, we can focus on the
one-particle sector of each operator
. Thus we only need to deal with
\begin{equation}\label{onept}
 e^{\theta_{\lambda,\gamma}(t)}= e^{\ii(\gamma-\lambda/2)q_{\rm tra}}e^{\ii\lambda
  q_{\rm tra}(t)}e^{-\ii(\gamma+\lambda/2)q_{\rm tra}},
\end{equation}
where $\theta_{\lambda,\gamma}(t),q_{\rm tra}$ and $q_{\rm tra}(t)$ are
one-particle operators corresponding to
$\Theta_{\lambda,\gamma}(t),Q_{\rm tra}$ and $Q_{\rm tra}(t)$. This fact has been
well appreciated in many works in the literature where only one species of
fermion appears, but it still holds in our situation where both fermion
and antifermion exist. This is because creation and annihilation
operators for fermions anticommute with those for antifermions. We will see
that the above expression can be computed exactly thanks to nice properties
of $q_{\rm tra}$ as explained below. 

In order to see how the one-particle operator \eqref{onept} acts on a one-particle state, we first recall that the one-particle Hilbert space of the Dirac theory is spanned by states $|\vp,c,s\rangle$ labeled by three-momentum $\vp$, charge $c$ and
spin $s$. Operators $Q_{\mathrm{L},\mathrm{R}}^+$ act on these states as
\begin{equation}
 Q_{\mathrm{L}}^{+}|\vp,c,s\rangle = \mathrm{sgn}(c)\theta(p^1)|\vp,c,s\rangle,\ \ Q_{\mathrm{R}}^+|\vp,c,s\rangle = \mathrm{sgn}(c)\theta(-p^1)|\vp,c,s\rangle.
\end{equation}
where $\mathrm{sgn}(x)$ and $\theta(x)$ are the sign function and the step
function respectively. Thus $|\vp,c,s\rangle$ are eigenstates of $Q_{\rm tra}^+=\frac{1}{2}(Q_{\mathrm{R}}^+-Q_{\mathrm{L}}^+)$
with eigenvalues $-\frac{1}{2}\mathrm{sgn}(c)\mathrm{sgn}(p^1)$, and
consequently $(Q_{\rm tra}^+)^2$ acts as $\frac{1}{4}$ on $|\vp,c,s\rangle$. Moreover the M\o ller operator $S_+=\lim_{t\to
\infty}e^{-\ii tH}e^{\ii tH_0}$ that intertwines $Q_{\rm tra}$ and $Q_{\rm tra}^+$ as
$S_+^{-1}Q_{\rm tra}^+S_+=Q_{\rm tra}$ allows
$Q_{\rm tra}^2$ to act on $|\vp,c,s\rangle$ in the same manner as $(Q_{\rm tra}^+)^2$:
$Q_{\rm tra}^2=S_+^{-1}(Q_{\rm tra}^+)^2S_+=\frac{1}{4}$. This follows from the fact that the M\o ller operator preserves the one-particle space, i.e.
\begin{align}
(Q_{\rm tra})^2|\vp,c,s\rangle &=S_+^{-1}(Q_{\rm tra}^+)^2\sum_{\vec{q},c',s'}|\vec{q},c',s'\rangle\langle \vec{q},c',s'|S_+|\vp,c,s\rangle \n
	&=\frac{1}{4}\sum_{\vec{q},c',s'}|\vec{q},c',s'\rangle\langle \vec{q},c',s'|S_+|\vp,c,s\rangle \n
    &=\frac{1}{4}|\vp,c,s\rangle,
\end{align}
where the summation over $\vec{q}$ is understood as an integration over $\vec{q}$ with appropriate normalization. This is best seen by representing the one-particle space as a space spanned by $\psi^\dagger_1|0\rangle$ and $\psi^\dagger_2|0\rangle$, where $\psi_1^\dagger(\vec{x}) = \int Dp\sum_s
 \adps u^{\dagger s}(p)e^{\ii p\cdot x}$ and $\psi_2^\dagger(\vec{x}) = \int Dp\sum_s
 \bdps v^{\dagger s}(p)e^{-\ii p\cdot x}$. Since $\psi^\dagger_i$ is a four-components spinor, the basis consists of 8 states that are not linearly-independent but can span the one-particle space. Both $H$ and $H_0$ act on this basis diagonally, thus the M\o ller operator preserves the space. Notice that this is also the
case for $Q_{\rm tra}(t)^2$. In what follows we shall denote one-particle
operators associated to $\Sigma(t)=Q_{\rm tra}(t)-Q_{\rm tra}(0)$, to
$[Q_{\rm tra},\Sigma(t)]$ and to $[Q_{\rm tra},[Q_{\rm tra},\Sigma(t)]]$ as
$\sigma(t), \sigma^{\prime}(t)$ and $\sigma^{\prime \prime}(t)$ respectively.\\
\ Following \cite{BD1}, those properties enable us to obtain
\begin{equation}
 e^{\theta_{\lambda,\gamma}(t)} = 1+\ii\sin \lambda
  \sigma(t)-2\sin^2\frac{\lambda}{2}\sigma(t)^2+2\sin\frac{\lambda}{2}\sin
  \gamma
  \sigma^{\prime}(t)+4\ii\sin\frac{\lambda}{2}\sin\Bigl(\frac{\lambda}{4}-\frac{\gamma}{2}\Bigr)\sin\Bigl(\frac{\lambda}{4}+\frac{\gamma}{2}\Bigr)\sigma^{\prime
  \prime}(t). \label{eq:5}
\end{equation}

\subsection{One-particle matrix elements}
\begin{figure}[!h]
\centering
\includegraphics[width=12cm]{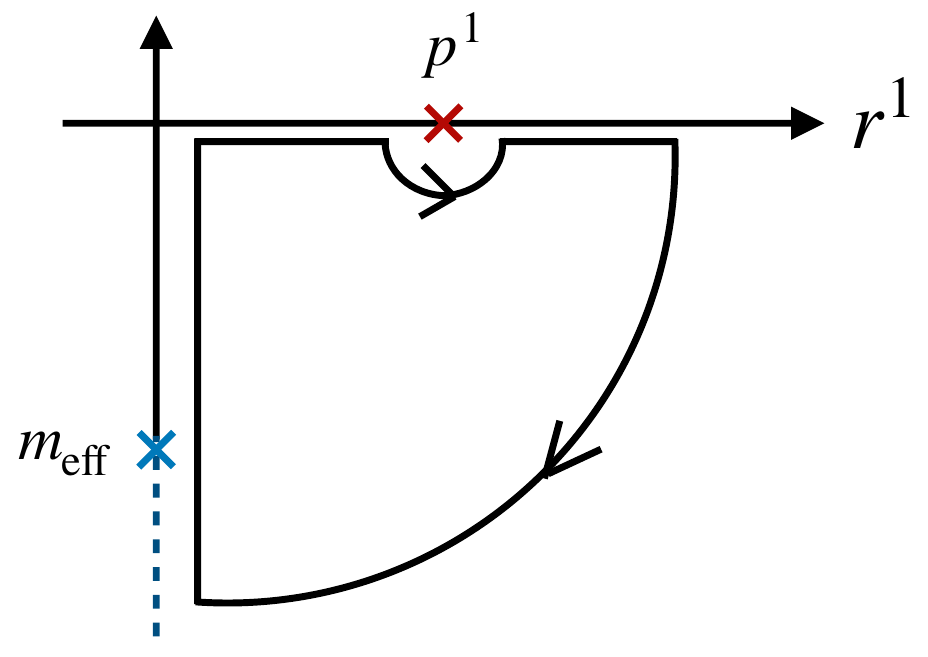}
\caption{The contour used to evaluate \eqref{eq:osc} with clockwise orientation, avoiding the pole at $r^1=p^1$. The dotted blue line shows the brunch cut.}
\label{fig:1}
\end{figure}

In this section we compute matrix elements of those
one-particle operators as they are essential in evaluating the
generating function. We first note that $\Sigma(t)$ can be expressed
as follows:
\begin{equation}
 \Sigma(t)=\ii\int_0^tdt^{\prime}[H,Q_{\rm tra}](t^{\prime})=\int
  d\tilde{x}\int_0^tdt^{\prime}\mathcal{J}^1(0,\tilde{x},t^{\prime}),
\end{equation}
where $\tilde{x}=(x^2,x^3)$ and $\mathcal{J}^1(\vx)=\bar{\psi}(\vx)\gamma^1\psi(\vx)$ is a $U(1)$ current along
the $x^1$ direction. In terms of mode operators, as long as it is
evaluated on the one-particle sector, it becomes
\begin{equation}
 \Sigma(t)=\int d^2\tilde{x}\int DpDq \sum_{s,r}(\adps \aqr
  \bar{u}^s(p)\gamma^1u^r(q)-\bdps \bqr
  \bar{v}^s(q)\gamma^1v^r(p))e^{-\ii(\tilde{p}-\tilde{q})\cdot
  \tilde{x}},
\end{equation}
where $\tilde{p}=(p^2,p^3)$ and $\tilde{q}=(q^2,q^3)$. The resulting matrix element for
$\Sigma(t)$, i.e. for $\sigma(t)$ is then 
\begin{equation}
\langle \vp,c^{\prime},s^{\prime}|\sigma(t)|\vq, c,s\rangle=
 \begin{cases}
  \Us \deltas \delta_{c,c^{\prime}} & \text{$c=+1$} \\
  -\Vs \deltas \delta_{c,c^{\prime}}& \text{$c=-1$}
 \end{cases},
\end{equation}
where we defined\footnote{Notice the difference between the definition of $\deltasss$ and $\deltas$} $\deltas=(2\pi)^3\delta^{(2)}(\tilde{p}-\tilde{q})\delta_t(E_{\vp}-E_{\vq})$ with
\begin{equation}
 \delta_t(p)=\frac{e^{\ii pt}-1}{2\pi \ii p}.
\end{equation}
Next we compute $\langle \vp,c^{\prime},s^{\prime}|\sigma^{\prime}(t)|\vq,
c,s\rangle=\langle \vp,c^{\prime},s^{\prime}|[Q_{\rm tra},\Sigma(t)]|\vq,
c,s\rangle=\langle \vp,c^{\prime},s^{\prime}|[Q_{\rm tra},Q_{\rm tra}(t)]|\vq,
c,s\rangle$. Noticing that the equation of motion
$d\Sigma(t)/dt=\ii[H,Q_{\rm tra}(t)]$ gives
\begin{equation}
\langle \vp,c^{\prime},s^{\prime}|Q_{\rm tra}(t)|\vp,
c,s\rangle=\frac{1}{\ii(\Ep-\Eq)}\frac{d}{dt}\langle\vp,c^{\prime},s^{\prime}|\sigma(t)|\vq,
c,s\rangle 
\end{equation}
we can expand $\langle \vp,c^{\prime},s^{\prime}|\sigma^{\prime}(t)|\vq,
c,s\rangle$ as
\begin{align}
 \langle \vp,c^{\prime},s^{\prime}|\sigma^{\prime}(t)|\vq,
c,s\rangle&=\sum_{c'',s''}\int Dr\Big[ \langle \vp,c^{\prime},s^{\prime}|Q_{\rm tra}|\vec{r},c'',s''\rangle\langle\vec{r},c'',s''|Q_{\rm tra}(t)|\vq,
c,s\rangle\nonumber \\
&\quad- \langle \vp,c^{\prime},s^{\prime}|Q_{\rm tra}(t)|\vec{r},c'',s''\rangle\langle\vec{r},c'',s''|Q_{\rm tra}|\vq,
c,s\rangle\Big]\nonumber\\
&=-\int
 \frac{dr^1}{2\pi2\Er}\frac{\bar{u}^{s^{\prime}}(p)\gamma^1(r_{\mu}\gamma^{\mu}+m)\gamma^1u^s(q)\delta_{c,1}+\bar{v}^{s^{\prime}}(q)\gamma^1(r_{\mu}\gamma^{\mu}-m)\gamma^1v^s(p^{\prime})\delta_{c,-1}}{(\Er-\Ep)(\Er-\Eq)}
 \nonumber \\
 &\quad \left.\times (e^{\ii t(\Ep-\Er)}-e^{\ii t(\Er-\Eq)})(2\pi)^2(\tilde{p}-\tilde{q})\delta_{c,c^{\prime}}\right|_{\tilde{r}=\tilde{p}=\tilde{q}}. \label{eq:1}
\end{align}
The equality $\tilde{r}=\tilde{p}=\tilde{q}$ will be implied for
 integrals over $r^1$ henceforth. The numerator of this expression is in fact simplified using the Dirac
algebra:
\begin{align}
 \bar{u}^{s^{\prime}}(p)\gamma^1(r_{\mu}\gamma^{\mu}+m)\gamma^1u^s(q)&=(\Er-\Eq)\bar{u}^{s^{\prime}}(p)u^s(q)+(r^1+q^1)\bar{u}^{s^{\prime}}(p)\gamma^1u^s(q) \label{eq:2}\\
 \bar{v}^{s^{\prime}}(q)\gamma^1(r_{\mu}\gamma^{\mu}+m)\gamma^1v^s(p)
 &=(\Er-\Ep)\bar{v}^{s^{\prime}}(q)v^s(p)+(r^1+p^1)\bar{v}^{s^{\prime}}(q)\gamma^1v^s(p).
\end{align}
Analogously, we also have
\begin{align}\label{dalgebra2}
 (\Ep-\Eq)\bar{u}^{s^{\prime}}(p)u^s(q)&=(p^
 1-q^1)\bar{u}^{s^{\prime}}(p)\gamma^1u^s(q)\\
 (\Ep-\Eq)\bar{v}^{s^{\prime}}(q)v^s(p)&=(p^1-q^1)\bar{v}^{s^{\prime}}(q)\gamma^1v^s(p).
\end{align}
Putting these relations into (\ref{eq:1}) then gives
\begin{align}\label{sigmap}
\langle \vp,c^{\prime},s^{\prime}|\sigma^{\prime}(t)|\vq,
c,s\rangle
&=A_1(l^{\prime};l)+A_2(l^{\prime};l)+\tilde{A}_1(l^{\prime};l)+\tilde{A}_2(l^{\prime};l)\n
&\quad+B_1(l^{\prime};l)+B_2(l^{\prime};l)+\tilde{B}_1(l^{\prime};l)+\tilde{B}_2(l^{\prime};l),
\end{align}
where $l$ and $l^{\prime}$ label triplets $(\vq,
c,s)$ and $(\vp,c^{\prime},s^{\prime})$ respectively, and 
\begin{align}
 A_1(l^{\prime};l)&=-\bar{u}^{s^{\prime}}(p)u^s(q)\int \frac{dr^1}{2\pi2\Er}\frac{e^{\ii t(\Ep-\Er)}}{\Er-\Ep}(2\pi)^2\delta^{(2)}(\tilde{p}-\tilde{q})\delta_{c,c^{\prime}}\delta_{c,1}
 \nonumber \\
 A_2(l^{\prime};l)&=-\bar{u}^{s^{\prime}}(p)\gamma^1u^s(q)\int \frac{dr^1}{2\pi2\Er}\frac{(r^1+q^1)e^{\ii t(\Ep-\Er)}}{(\Er-\Ep)(\Er-\Eq)}(2\pi)^2\delta^{(2)}(\tilde{p}-\tilde{q})\delta_{c,c^{\prime}}\delta_{c,1}
 \nonumber \\
 \tilde{A}_1(l^{\prime};l)&=-\bar{u}^{s^{\prime}}(p)u^s(q)\int \frac{dr^1}{2\pi2\Er}\frac{e^{-\ii t(\Eq-\Er)}}{\Er-\Ep}(2\pi)^2\delta^{(2)}(\tilde{p}-\tilde{q})\delta_{c,c^{\prime}}\delta_{c,1}
 \nonumber \\
 \tilde{A}_2(l^{\prime};l)&=-\bar{u}^{s^{\prime}}(p)\gamma^1u^s(q)\int \frac{dr^1}{2\pi2\Er}\frac{(r^1+q^1)e^{-\ii t(\Eq-\Er)}}{(\Er-\Ep)(\Er-\Eq)}(2\pi)^2\delta^{(2)}(\tilde{p}-\tilde{q})\delta_{c,c^{\prime}}\delta_{c,1}
 \nonumber \\
B_1(l^{\prime};l)&=-\bar{v}^{s^{\prime}}(q)v^s(p)\int \frac{dr^1}{2\pi2\Er}\frac{e^{\ii t(\Ep-\Er)}}{\Er-\Eq}(2\pi)^2\delta^{(2)}(\tilde{p}-\tilde{q})\delta_{c,c^{\prime}}\delta_{c,-1}
 \nonumber \\
B_2(l^{\prime};l)&=-\bar{v}^{s^{\prime}}(q)\gamma^1v^s(p)\int
 \frac{dr^1}{2\pi2\Er}\frac{(r^1+p^1)e^{\ii t(\Ep-\Er)}}{(\Er-\Eq)(\Er-\Ep)}(2\pi)^2\delta^{(2)}(\tilde{p}-\tilde{q})\delta_{c,c^{\prime}}\delta_{c,-1}.\nonumber \\
\tilde{B}_1(l^{\prime};l)&=-\bar{v}^{s^{\prime}}(q)v^s(p)\int \frac{dr^1}{2\pi2\Er}\frac{e^{-\ii t(\Eq-\Er)}}{\Er-\Eq}(2\pi)^2\delta^{(2)}(\tilde{p}-\tilde{q})\delta_{c,c^{\prime}}\delta_{c,-1}
 \nonumber \\
\tilde{B}_2(l^{\prime};l)&=-\bar{v}^{s^{\prime}}(q)\gamma^1v^s(p)\int
 \frac{dr^1}{2\pi2\Er}\frac{(r^1+p^1)e^{-\ii t(\Eq-\Er)}}{(\Er-\Eq)(\Er-\Ep)}(2\pi)^2\delta^{(2)}(\tilde{p}-\tilde{q})\delta_{c,c^{\prime}}\delta_{c,-1}.
\end{align}
Firstly let us deal with $A_1(l^{\prime};l)$. We need to perform the integral
only over $r^1\geq 0$ as the integral is even under $r^1
\mapsto -r^1$. Furthermore the integrand has branch
cuts on the imaginary axis starting at $\pm \ii\me=\pm \ii\sqrt{\vec{r}^2+m^2}$ \footnote{The $\vec{r}$ dependence will be implied henceforth
unless otherwise stated.}. Here we can interpret $\me$ as an effective mass taking
account for the transverse momenta. Thus we
deform the contour towards the negative imaginary direction so that it
vanishes under the large $t$ limit (see Fig.\ref{fig:1}), obtaining

\begin{equation}
 2\int_0^{\infty}
  \frac{dr^1}{2\pi2\Er}\frac{e^{\ii t(\Ep-\Er)}}{\Er-\Ep}
  = -\frac{\ii}{2|p^1|}-
  \frac{\ii}{\pi}\int_0^{\me}\frac{d\rho}{\sqrt{\me^2-\rho^2}}\frac{e^{\ii t(\Epp-\sqrt{\me^2-\rho^2})}}{\sqrt{\me^2-\rho^2}-\Ep}+\mathcal{O}(e^{-mt}). \label{eq:osc}
\end{equation}
Similarly $A_2(l^{\prime};l)$ can be
calculated as
\begin{align}
 2p^1\int_0^{\infty}
 \frac{dr^1}{2\pi2\Er}\frac{e^{\ii t(\Ep-\Er)}}{(\Er-\Ep)(\Er-\Eq)}&=
\frac{\ii q^1}{2(\Ep-\Eq)}\Bigl(\frac{1}{|q^1|}e^{\ii t(\Ep-\Eq)}-\frac{1}{|p^1|}\Bigr) \nonumber \\
&\quad-\frac{\ii p^1}{\pi}\int_0^{\me}\frac{d\rho}{\sqrt{\me^2-\rho^2}}\frac{e^{\ii t(\sqrt{\me^2-\rho^2}-\Eq)}}{(\sqrt{\me^2-\rho^2}-\Ep)(\sqrt{\me^2-\rho^2}-\Eq)} \nonumber \\
 &\quad+\mathcal{O}(e^{-mt}).
\end{align}
With the help of the relation \eqref{dalgebra2}, combining everything together yields
\begin{equation}
 A_1(l^{\prime};l)+A_2(l^{\prime};l)=\mathrm{oscillatory\
  terms}+
\begin{cases}
 -\pi\Us \delta_t(\Ep-\Eq)\delta_{c,1} &\text{$p^1,q^1>0$} \\
 -\pi\Us \delta^{\prime}_t(\Ep-\Eq)\delta_{c,1}& \text{$p^1<0<q^1$} \\
\pi\Us \delta^{\prime}_t(\Ep-\Eq)\delta_{c,1}& \text{$q^1<0<p^1$} \\
\pi\Us \delta_t(\Ep-\Eq)\delta_{c,1} &\text{$p^1,q^1<0$}
\end{cases},
\end{equation}
with $\delta^{\prime}_t(p)=(e^{\ii tp}+1)/2\pi\ii p$. The oscillatory terms in fact vanish as $t$ becomes large: see
Appendix \ref{osci}. In the same manner we also find $\tilde{A}_1(l^{\prime};l)+\tilde{A}_2(l^{\prime};l)=-A_1(l^{\prime};l)-A_2(l^{\prime};l)$, implying $A_1(l^{\prime};l)+A_2(l^{\prime};l)+\tilde{A}_1(l^{\prime};l)+\tilde{A}_2(l^{\prime};l)=0$. The same story holds for $B$'s as well, namely
\begin{equation}
 B_1(l^{\prime};l)+B_2(l^{\prime};l)=\mathrm{oscillatory\
  terms}+
\begin{cases}
 \pi\Vs \delta_t(\Ep-\Eq)\delta_{c,1} &\text{$p^1,q^1>0$} \\
 \pi\Vs \delta^{\prime}_t(\Ep-\Eq)\delta_{c,1}& \text{$p^1<0<q^1$} \\
-\pi\Vs \delta^{\prime}_t(\Ep-\Eq)\delta_{c,1}& \text{$q^1<0<p^1$} \\
-\pi\Vs \delta_t(\Ep-\Eq)\delta_{c,1} &\text{$p^1,q^1<0$}
\end{cases},
\end{equation}
and
$B_1(l^{\prime};l)+B_2(l^{\prime};l)+\tilde{B}_1(l^{\prime};l)+\tilde{B}_2(l^{\prime};l)=0$. Hence merging everything together, upon taking the large $t$ limit, we obtain
\begin{equation}
 \langle \vp,c^{\prime},s^{\prime}|\sigma^{\prime}(t)|\vq,
c,s\rangle=0,
\end{equation}
and automatically $\langle \vp,c^{\prime},s^{\prime}|\sigma^{\prime \prime}(t)|\vq,
c,s\rangle=0$. Moreover utilizing the building blocks we computed above,
 we find
\begin{align}
 \langle \vp,c^{\prime},s^{\prime}|\sigma(t)^2|\vq,
 c,s\rangle
 &= \begin{cases}
  \Us W_{\vp,\vq} & \text{$c=+1$} \\
  \Vs W_{\vp,\vq}& \text{$c=-1$}
 \end{cases},
\end{align}
where
\begin{equation}
 W_{\vp,\vq}=\begin{cases}
			        \deltas &\text{$p^1,q^1>0$} \\
			       \deltass& \text{$p^1<0<q^1$} \\
-\deltass& \text{$q^1<0<p^1$} \\
-\deltas &\text{$p^1,q^1<0$}
			      \end{cases},
\end{equation}
with $\deltass=(2\pi)^3\delta^{(2)}(\tilde{p}^{\prime}-\tilde{q})\delta^{\prime}_t(\Ep-\Eq)$.
We now put everything into (\ref{eq:5}) and obtain, for fermions ($c=1$)
\begin{align}
 \langle \vp,c^{\prime},s^{\prime}|e^{\theta_{\lambda,\gamma}(t)}-1|\vq, c,s\rangle &= \begin{cases}
  (e^{\ii\lambda}-1)\Us
		     \deltas \delta_{c^{\prime},c}&\text{$p^1,q^1>0$} \\
  (e^{\ii\lambda}-1)\Us \deltass \delta_{c^{\prime},c}&\text{$p^1<0<q^1$} \\
  (1-e^{-\ii\lambda})\Us \deltass \delta_{c^{\prime},c}& \text{$q^1<0<p^1$} \\
  (1-e^{-\ii\lambda})\Us \deltas \delta_{c^{\prime},c}& \text{$p^1,q^1<0$},
 \end{cases}
\end{align}
and for antifermions ($c=-1$)
\begin{align}
 \langle \vp,c^{\prime},s^{\prime}|e^{\theta_{\lambda,\gamma}(t)}-1|\vq,c,s\rangle &= \begin{cases}
  (e^{-\ii\lambda}-1)\Vs \deltas \delta_{c^{\prime},c}& \text{$p^1,q^1>0$} \\
  (e^{-\ii\lambda}-1)\Vs \deltass \delta_{c^{\prime},c}& \text{$p^1<0<q^1$} \\
  (1-e^{\ii\lambda})\Vs \deltass \delta_{c^{\prime},c}& \text{$q^1<0<p^1$} \\
  (1-e^{\ii\lambda})\Vs \deltas \delta_{c^{\prime},c}& \text{$p^1,q^1<0$}.
 \end{cases}
\end{align}

\subsection{Levitov-Lesovik formula in 3+1 dimensions}
The quantity we are initially interested in is the long-time limit of
the cumulant generating function (\ref{eq:gf}). By means of Klich's
trace formula \cite{klich}, it was shown in \cite{BD1} that the generating function can be 
expressed in the following form:
\begin{equation}
 \mathrm{Tr}(\rho_{\mathrm{s}}e^{\Theta_{\lambda,\gamma}(t)}) =
  \det\bigl(1+n_{\mathrm{s}}(e^{\theta_{\lambda,\gamma}(t)}-1)\bigr),
\end{equation}
where
\begin{equation}
 \langle \vp,c^{\prime},s^{\prime}|n_{\mathrm{s}}|\vq, c,s\rangle = \begin{cases}
		   n_+(\vp)2\Ep(2\pi)^3\delta^{(3)}(\vp-\vq)\delta_{c^{\prime}c}\delta_{s,s'} & \text{$c=1$} \\
		    n_-(\vp)2\Ep(2\pi)^3\delta^{(3)}(\vp-\vq)\delta_{c^{\prime}c}\delta_{s,s'} & \text{$c=-1$}
		   \end{cases}
\end{equation}
with $n_{\pm}(\vp)=1/(e^{V_{\pm}(\vp)}-1)$. Note that the trace on the LHS
is done over the full Hilbert space, while the determinant on the RHS is
over the one-particle Hilbert space.\\
\ It can be readily shown\footnote{For a detailed derivation, see
appendix \ref{asymp}.} that for $t\to \infty$, contributions from the
diagonal matrix elements become dominant, and we find
\begin{align}
 \log\mathrm{Tr}(\rho_{\mathrm{s}}e^{\Theta_{\lambda,\gamma}(t)})&=2tL^2\sum_{\epsilon=\pm}\int\frac{d^2{\tilde{p}}}{(2\pi)^2} \n
 &\quad\times\int_{\me(\tilde{p})}^{\infty}\frac{dE}{2\pi}\log\bigl[1+n_{\epsilon,\mathrm{L}}(n_{\epsilon,\mathrm{R}}-1)(1-e^{\ii\epsilon \lambda})+n_{\epsilon,\mathrm{R}}(n_{\epsilon,\mathrm{L}}-1)(1-e^{-\ii\epsilon \lambda})\bigr],
\end{align}
where $n_{\epsilon;\mathrm{L},\mathrm{R}}(E)$ are fermionic filling functions defined as
\begin{equation}
n_{\epsilon;\mathrm{L},\mathrm{R}}(E)=\frac{1}{e^{\beta_{\mathrm{L},\mathrm{R}}(E-\epsilon \mu_{\mathrm{L},\mathrm{R}})}+1}.
\end{equation}
Hence the desired scaled cumulant generating function is
\begin{align}
F(\lambda)&=2\sum_{\epsilon=\pm}\int\frac{d^2{\tilde{p}}}{(2\pi)^2}\int_{\me(\tilde{p})}^{\infty}\frac{dE}{2\pi}\log\bigl[1+n_{\epsilon,\mathrm{L}}(n_{\epsilon,\mathrm{R}}-1)(1-e^{\ii\epsilon
  \lambda})+n_{\epsilon,\mathrm{R}}(n_{\epsilon,\mathrm{L}}-1)(1-e^{-\ii\epsilon
  \lambda})\bigr]. \nonumber \\
 &:= f(\lambda;\betal, \betal\mul)+f(-\lambda;\betar,\betar\mur), \label{fac}
\end{align}
where $f(\lambda;\beta,\beta\mu)$ is given by
\begin{equation}
 f(\lambda
;\beta,\beta\mu)=2\sum_{\epsilon=\pm}\int\frac{d^2{\tilde{p}}}{(2\pi)^2}\int_{\me(\tilde{p})}^{\infty}\frac{dE}{2\pi}\log\bigl[1+n_{\epsilon}(e^{\ii\epsilon
  \lambda}-1)\bigr]. \label{eq:ll}
\end{equation}

Some comments are in order. We first notice that this can be seen as a
simple relativistic generalization of the celebrated Levitov-Lesovik formula
\cite{LL1} in three spatial
dimensions. The
two possible spins and charges are responsible for a prefactor 2 and a
sum over two signs $\epsilon=\pm$ respectively. Note that all transmission
coefficients are 1 as this is a free model without impurities.  Thanks to this fact the
SCGF factorizes as \eqref{fac}, and consequently enables us to
have a clear interpretation of it: the SCGF associated with the charge
transport is a sum of independent Bernoulli processes. Recall that
the Bernoulli process is a time-discrete stochastic process whose possible
events at each
frame (trial of a jump) are only either success or failure of the
jump. For instance a
fermion that was initially prepared in the left (resp. right) subsystem
with energy $E$ and charge 1 (the value of spin does not matter) is
transferred from left to right (resp. from right to left) with a success
probability $n_{+,\mathrm{L}}(E)$ (resp. $n_{+,\mathrm{R}}(E)$), thus has a
SCGF  $\log[1+n_{+,\mathrm{L}}(E)(e^{\ii\lambda}-1)]$
(resp. $\log[1+n_{+,\mathrm{R}}(E)(e^{-\ii\lambda}-1)]$). The number of
fermions that jump successfully during time $t$ has then a Binomial
distribution $B(n_t,p)$ where $n_t$ is the number of trials during time
$t$, and the probability of obtaining $k$ successes $P(k;n_t,p)$ is given by
\begin{equation}
 P(k;n_t,p)=\begin{pmatrix}
	 n_t\\
	       p
	      \end{pmatrix}p^k(1-p)^{n_t-k}.
\end{equation}
Since the CGF of independent
processes are additive, we can obtain the SCGF \eqref{fac} by integrating and summing over energy and
charge. This interpretation for the charge transport in the Dirac theory
would hold in any dimension.\footnote{It should be stressed that
one should not confuse this situation with the low temperature limit
$\beta \mu \gg 1$ of the original Levitov-Lesovik formula
\cite{LL1,LL2}. Although the authors of \cite{LL2} considered the probability
distribution as a Binomial distribution under that limit, this and our interpretation have twofold
differences. First, they are dealing with a model with a generic
transmission coefficient $0<\mathcal{T}<1$ while ours is free ($\mathcal{T}=1$). Furthermore
they take the limit of negligible temperature where the CGF does not depend on the energy of electrons anymore, and argue that the resulting
distribution is Binomial with $\mathcal{T}$ being the (energy-independent) success probability. This is essentially different from
our interpretation that the occupation function, rather than $\mathcal{T}$, plays a role of the
(success) probability of each jump. In the case of free theory, the low temperature limit simply results in perfect
transmission, i.e. transfer without thermal fluctuations.} It is illuminating to contrast this Bernoullian interpretation for the charge
transport with the Poissonian interpretation for the energy transport in higher dimensional free models and 2D conformal field theories \cite{BDreview,BD2012,doyonKG}, which indicates that energy quanta and charge quanta are transferred according to different statistics in those systems.
\  We also observe that when the two
temperatures are equal $T_{\mathrm{L}}=T_{\mathrm{R}}=T$, this
formula satisfies the fluctuation theorem:
$F(\lambda)=F(-\lambda+\ii(\mul-\mur)/T)$ \cite{EHM}. This is
also evident as a consequence of the time-reversal symmetry of the Dirac
theory \cite{BD3}. Performing
the integral, we can derive the analytic expression (see Appendix \ref{integ}) of the chiral SCGF
(\ref{eq:ll}) which reads
 \begin{subnumcases}{f(\lambda;\beta,\beta\mu)=}
		       -\sum_{\epsilon=\pm}\sum_{n=1}^{\infty}\frac{(-1)^n(1+\beta mn)}{2\pi^2\beta^3n^4}e^{-\beta
		       n(m-\epsilon \mu)}(e^{\ii\epsilon \lambda
		       n}-1)&
		       $|\mu|<m$ \label{chiralfcs1} \\
		       f_0(\lambda;\beta,\mu)+f_{\mathrm{M}}(\lambda;\beta,\mu)
		       & $m<\mu$ \label{chiralfcs2}\\
		       f_0(\lambda;\beta,\mu)+f_{\mathrm{M}}(-\lambda;\beta,-\mu)
		       & $m<-\mu$, \label{chiralfcs3}
		      \end{subnumcases}
where we defined $f_0(\lambda,\beta,\mu)$, which is the massless limit of
(\ref{eq:ll}), and $f_{\mathrm{M}}(\lambda;\beta,\mu)$ as
\begin{align}
  f_0(\lambda;\beta,\beta\mu)&=\frac{\ii\mu^3\lambda}{12\pi^2}
-\frac{1}{2\pi^2\beta^3}\sum_{\epsilon=\pm}\sum_{n=1}^{\infty}\frac{(-1)^n}{n^4}\frac{1+(1+\epsilon
 \mu
 \beta n)^2}{2}(e^{\ii\epsilon \lambda n}-1)\label{eq:ll2}\\
f_{\mathrm{{M}}}(\lambda;\beta,\beta\mu)
 &=\frac{\ii m^2}{12\pi^2}(2m-3\mu)\lambda+\frac{1}{\pi^2\beta^3}\sum_{n=1}^{\infty}\frac{(-1)^n}{n^4}\biggl[\frac{\beta^2m^2n^2}{4}(e^{\ii\lambda
 n}-1) \nonumber \\
 &\ +\Bigl\{e^{-\beta \mu n}(\sinh(\beta mn)-\beta mn\cosh(\beta mn))+\frac{\beta^2m^2n^2}{4}\Bigr\}(e^{-\ii\lambda
 n}-1)
\biggr]. \label{eq:ll3}
\end{align}
That the SCGFs at $m<\mu$ and $m<-\mu$ have similar forms is due to the charge conjugation symmetry in the
Dirac theory. It is worth noting that the parity symmetry, a symmetry under the
swapping of left- and right-handed spinors, plays no role here since the distinction between the two spinors does not affect anything in the charge transport. Those SCGFs basically comprise
 (possibly) a linear term and a sum of Poisson-like terms with
 coefficients that are not necessarily positive. The linear term in
 fact represents a perfect transmission \cite{CSS}:
 perfect transmission means a transfer with the success
 probability 1 now, hence the corresponding SCGF is
 $\log(1+e^{\ii\lambda}-1)=\ii\lambda$ (modulo $2\pi$).\par
\ There is one more thing that deserves to be detailed. For ease of explanation we shall assume $\mu>0$ hereafter. Let us remind that as a
consequence of the charge quantization, our SCGF has a periodicity
$F(\lambda)=F(\lambda+2\pi)$ as do other CGFs associated with charge
transfers. It is then natural to ask how the SCGF looks like in the
fundamental domain $(-\pi,\pi)$. Further restricting our attention
to the case $m<\mu$, when $\lambda \in (-\pi,\pi)$, the series in (\ref{eq:ll2}) and (\ref{eq:ll3}) converge, yielding
\begin{align}
 f_0(\lambda;\beta,\beta\mu)&=\frac{1}{12\pi^2\beta^3}\Bigl[\frac{1}{4}(\beta\mu+\ii\lambda)^4-\frac{1}{4}(\beta\mu)^4+\frac{\pi^2}{2}(\beta\mu+\ii\lambda)^2-\frac{\pi^2}{2}(\beta\mu)^2\Bigr], \label{eq:ll21}\\
 f_{\mathrm{M}}(\lambda;\beta,\beta\mu)&=\frac{\ii m^3\lambda}{6\pi^2}-\frac{m^2}{8\pi^2\beta}\bigl[(\beta
 \mu +\ii\lambda)^2-(\beta \mu)^2\bigr] \nonumber \\
&\quad -\frac{1}{\pi^2\beta^3}\sum_{n=1}^{\infty}\frac{(-1)^n}{n^4}e^{-\beta \mu n}(\sinh(\beta mn)-\beta mn\cosh(\beta mn))(e^{-\ii\lambda
 n}-1).
\end{align}
In the massless case, the SCGF becomes a finite
polynomial of $\lambda$ by virtue of the presence of both fermions and
antifermions in this model. Therefore we have cumulants only up to
fourth order\footnote{We expect that, in ($d+1$)-D Dirac theory, cumulants are generically non-zero only up to ($d+1$)-th order.}. Notice, however, that this does not mean that the full
SCGF is also a finite polynomial - it is not analytic (not Taylor
expandable) in the entire domain. In fact when $m<\mu$, it is not even differentiable, which
amounts to a quantitatively different behavior of the charge current
depending on $m \gtrless \mu$ as we shall see below. 

\begin{figure}[!htb]
 \centering
\includegraphics[width=12cm, clip]{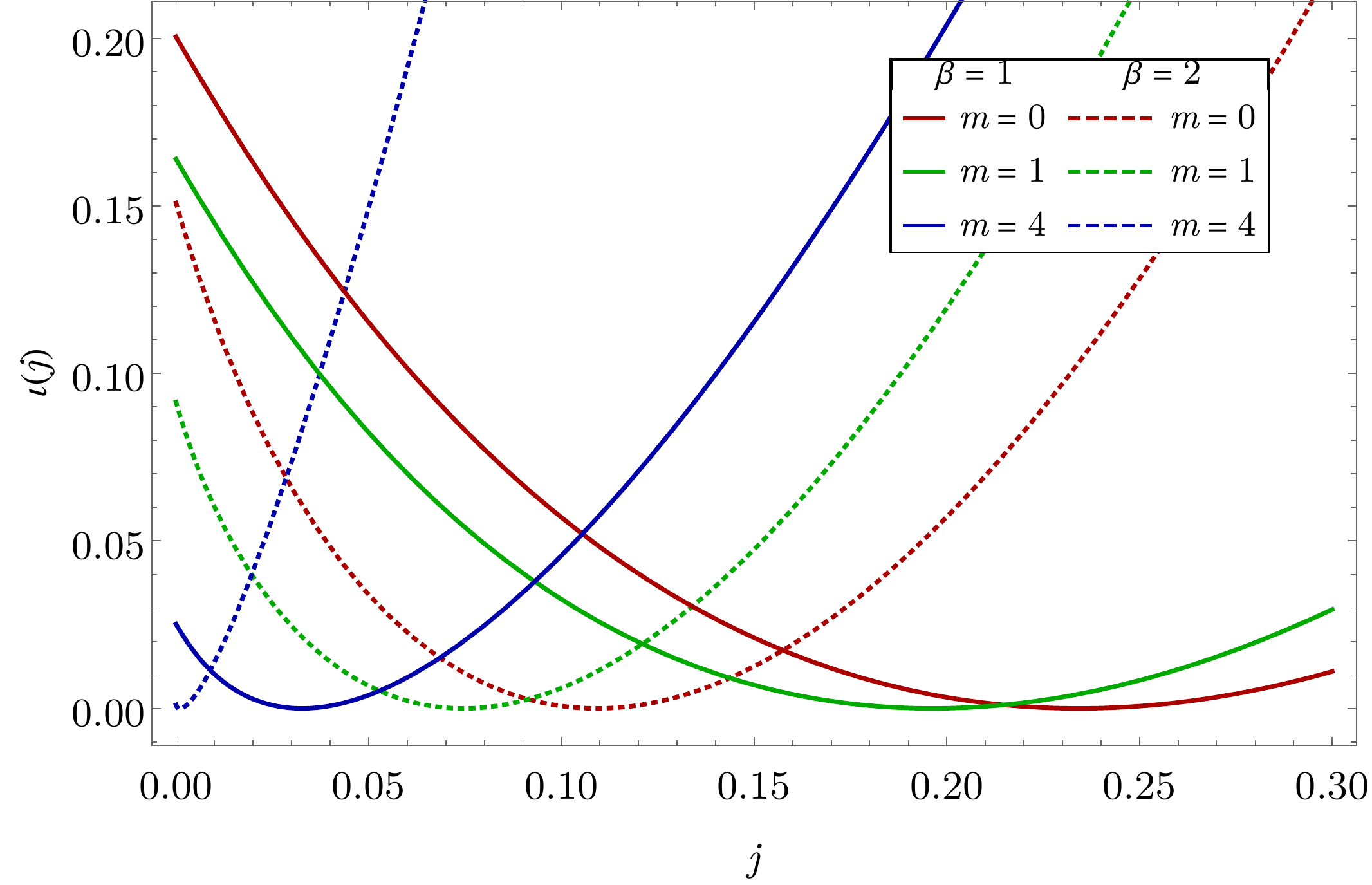}
\caption{The chiral LDF (CLDF) $\iota(j)$ is depicted with $\mu=2$ varying the mass $m$ and the inverse temperature $\beta$. The value of the mass is chosen so that the lines with different colors correspond to different regimes (red, green and blue lines are for massless $m=0$, small-mass $m<\mu$, and large-mass $m>\mu$ regimes respectively).  Solid lines
 correspond to the CLDF for $\beta=1$ while dashed lines describes that for $\beta=2$. As explained in the main text, $j$'s at which the CLDF takes its minimum is the value of the chiral NESS current $j=\langle j_Q\rangle$. We observe that as the temperature decreases, the chiral NESS current in the large-mass phase becomes distinctively small compared to those in other regimes. }
\label{fig:2}
\end{figure}

Before focusing on the charge current, let us introduce the notion of {\it large deviation function} (LDF) $I(J)$ defined as the Legendre transform of the SCGF \eqref{fac}
\begin{equation}\label{ldf}
I(J)=\max_{\lambda \in \mathbb{R}}(\lambda J-F(\lambda)).
\end{equation}
This function inherits the convexity of $F(\lambda)$ and takes its minimum, which is zero, when $J=\langle J_Q\rangle$ where $\langle J_Q\rangle$ is the NESS charge current. Thus the function $-I(J)$ bears a similarity with entropy in equilibrium that is maximized by the (generalized) thermal state. As seen in Fig.\ref{fig:2}, the chiral LDF $\iota(j)\equiv\max_{\lambda \in \mathbb{R}}(\lambda j-f(\lambda))$, hence the full LDF, indeed satisfies the above properties of generic LDFs.

Once we obtain the SCGF, its multiple differentiations with respect to $\lambda$ evaluated at $\lambda=0$ yield all the
cumulants. In particular, the average of the charge current in the NESS
 $\langle J_Q(\betal,\betal \mul;\betar,\betar
\mur)\rangle=\left. dF(\lambda)/d(\ii\lambda) \right|_{\lambda=0}$ is of great importance in actual experiments. The chiral part $\langle j_Q(\beta,\beta
\mu)\rangle=\left. df(\lambda;\beta,\beta
\mu)/d(\ii\lambda)\right|_{\lambda=0}$ reads,
\begin{equation}
\langle j_Q(\beta,\beta
\mu)\rangle=\begin{dcases*}
      \mathcal{O}(\beta^{-2}) & $\mu<m$ \\
  \frac{\mu^3}{12\pi^2}+\frac{m^3}{6\pi^2}-\frac{m^2\mu}{8\pi^2}+\frac{\mu}{12\beta^2}+\frac{m}{12\beta^2}+\mathcal{O}(\beta^{-2})
      & $m<\mu$.
 \end{dcases*}
\end{equation}
The phase $\mu<m$ can be considered as an {\it insulating} phase in the sense that the charge current is effectively zero in the low
 temperature limit: jumps made by
 particles are less probable because the occupation function $n(E)$,
 which is supposed to be the probability of each jump, can
 take a value only less than one half (see Fig.\ref{fig:2}). The existence of such phase is a peculiarity in Dirac theory which is forbidden in non-relativistic free fermionic systems (see Appendix \ref{lifshitz}). It is also of particular interest to observe its massless limit since most
 Dirac fermions that exist as low-energy theories in unconventional
 matters such as Dirac semimetals are
 massless. The
 current in the massless limit is given exactly by
\begin{equation}
\langle J_Q(\betal,\betal \mul;\betar,\betar
\mur)\rangle=
\frac{\mu^3_{\mathrm{L}}-\mu^3_{\mathrm{R}}+\pi^2(\mu_{\mathrm{L}}T^2_{\mathrm{L}}-\mu_{\mathrm{R}}T^2_{\mathrm{R}})}{12\pi^2}. \label{eq:charge}
\end{equation}
In the zero
temperature limit, this reduces to $\langle J_Q(T_{\mathrm{L},\mathrm{R}}=0)\rangle=(\mul^3-\mur^3)/12\pi^2$, which can be expected given the chiral separation \eqref{fac} and on the basis of dimensional analysis. We expect
that this is generalized to $\langle J_Q(T_{\mathrm{L},\mathrm{R}}=0)\rangle \propto
\mul^d-\mur^d$ in $d$ spatial dimensional Dirac models.

\subsection{Extended fluctuation relation}
It was advocated in \cite{BD3} that if systems whose
dynamics satisfy the pure tansmission condition
$S\tilde{Q}=-\tilde{Q}S$, the {\it extended fluctuation relation} (EFR) holds. Here
$S=S_-^{-1}S_+$ with $S_-=\lim_{t\to -\infty}e^{-\ii tH}e^{\ii tH_0}$ is the
$S$-matrix, and for the charge transport the EFR reads
\begin{equation}
 F(\lambda)=\int_0^{\ii\lambda}dz\langle J_Q(\betal, \betal \mul+z;\betar,\betar \mur-z)\rangle
\end{equation}
or alternatively, due to the factorization \eqref{fac}, 
\begin{equation}
 f(\lambda;\beta,\beta \mu)=\int_0^{\ii\lambda}dz\langle j_Q(\beta,\beta \mu+z)\rangle.
\end{equation}
It is immediate to see that the
Dirac theory indeed satisfies it: let us again focus on the massless
case where $\langle j_Q(\beta,\beta \mu)\rangle=(\beta^3\mu^3+\pi^2\beta\mu)/12\pi^2\beta^3$. Upon
shifting $\beta \mu \mapsto \beta \mu+z$ and an integral over $z$,
(\ref{eq:ll21}) is reproduced. Once we obtain (\ref{eq:ll21}), a
representation of the SCGF for the entire domain, (\ref{eq:ll2}), is computed as
a Fourier series. Thus we need only the average of the current to
acquire the SCGF (all other higher cumulants are unnecessary). Notice
that in the course of deriving Fourier coefficients, we need to figure
out terms associated with perfect transmission. In order to see this, we
can take a zero temperature limit $\beta \to \infty$ where thermal
fluctuation is negligible: remaining terms, in our case
$\ii\mu^3\lambda/12\pi^2$, are the sought ones. \par
\ Finally let us emphasize why EFR is useful. The approach we have
taken, which is based on the two-point von Neumann measurement, enables
us to directly compute the SCGF. In view of that approach, one might
find that the EFR is just a curious relation. However, when one knows the average
current first rather than having the SCGF first (for instance, this is the case in hydrodynamics), EFR becomes extremely
powerful: one can obtain the SCGF just by integrating the current with
shifting parameters properly. We demonstrate how we can gain the SCGF for the energy
transport in Appendix \ref{energy} as well as reporting the SCGF that combines both
charge and energy transports. Furthermore, for a comparison, we briefly
examine the charge transport in the Lifshitz fermions
(i.e. non-relativistic fermion system with a dispersion relation
$E=|p|^{z}/2m$) in Appendix \ref{lifshitz}. 

\section{Discussion and conclusion}
In this manuscript we have studied the non-equilibrium $U(1)$ charge
transport of the Dirac theory. We derived the
NESS density operator in the same spirit as \cite{doyonKG}, and computed the SCGF associated with the charge
transport. The so-determined SCGF was then interpreted as an extension of the
Levitov-Lesovik formula to higher dimensions. Remarkably, in the
massless limit, cumulants of the SCGF exist up
to fourth order. We expect that, generically, in $D$ space-time dimensions,
the cumulants are nonzero only up to $D$-th order. We also found an insulating regime, which does not exist in the non-relativistic free fermion systems, where the NESS current is negligibly small for small temperatures. Finally the validity of the EFR for
the {\it charge} transfer was confirmed. One of natural
extensions of our results would be to consider an impurity at the junction $x^1=0$ \cite{BDV,bertini2}, giving rise to the transmission coefficient $T(p^1)$ that is not unity. In this case, we still expect that the chiral SCGF takes a similar form as (\ref{eq:ll}):
\begin{equation}
 f(\lambda
;\beta,\beta\mu)=2\sum_{\epsilon=\pm}\int\frac{d^2{\tilde{p}}}{(2\pi)^2}\int_{\me(\tilde{p})}^{\infty}\frac{dE}{2\pi}\log\bigl[1+Tn_{\epsilon}(e^{\ii\epsilon
  \lambda}-1)\bigr].
\end{equation}
This would be straightforwardly derived by writing down the NESS density operator with taking accounting of the impurity \cite{mintchev1}.

Another possible generalization could be extending to generic spacial dimensions
$d$. To do so, one should be aware of the different nature of the spinor
representation of $SO(d,1)$ in even and odd dimensions: in odd spatial
dimensions there exist Weyl spinors, but this is not the case in even
spatial dimensions. It would be of course interesting to study the effect of interactions, but for that purpose, our approach might not be the most efficient way. Instead, focusing on the long-wavelength physics, one can study the dynamics of the Dirac fermions as the {\it Dirac fluid} \cite{LDS,LF}.
\section{Acknowledgements}
I thank Benjamin Doyon and Joe Bhaseen for useful discussions, and acknowledge the support of
the Takenaka scholarship foundation. This work is also partly supported by the ERC advanced grant NuQFT.
\appendix

\section{Time-evolution in B-representation}\label{bconv}
In this section, we explicitly demonstrate that $\psi_{\rm B}(\vx,t)=\lim_{t\to\infty}S(\psi_{\rm A}(\vx,t))$  \cite{doyonKG}. By construction, it is immediate to see that
\begin{equation}\label{b-evolve}
\psi_{{\rm B},a}(\vx,t)=\sum_b\int d^3y\{\psi_{{\rm A},a}(\vx,t),\psi^\dagger_{{\rm A},b}(\vec{y})\}\psi_{{\rm B},b}(\vec{y})
\end{equation}
solves the equation motion of the Dirac theory $(i\gamma^\mu\partial_\mu-m)\psi_{{\rm B},a}(\vx,t)=0$. The anticommutator $\{\psi_{{\rm A},a}(\vx,t),\psi^\dagger_{{\rm A},b}(\vec{y})\}$ is readily evaluated by the direct computation:
\begin{align}
\{\psi_{{\rm A},a}(\vx,t),\psi^\dagger_{{\rm A},b}(\vec{y})\}&=\int \frac{d^3p}{(2\pi)^32\Ep}\sum_s\bigl[u^s_a(p)u^{s\dagger}_b(p)e^{-\ii\Ep t}+v^s_a(\bar{p})v^{s\dagger}_b(\bar{p})e^{\ii\Ep t}\bigr]e^{\ii\vp\cdot(\vx-\vec{y})} \n
	&=\int \frac{d^3p}{(2\pi)^32\Ep}\sum_s\mathbb{M}_{ab}^s(\vp,t)e^{\ii\vp\cdot(\vx-\vec{y})},
\end{align}
where $\bar{p}=(p^0,-\vp)$ and $\mathbb{M}_{ab}^s(\vp,t):=u^s_a(p)u^{s\dagger}_b(p)e^{-\ii\Ep t}+v^s_a(\bar{p})v^{s\dagger}_b(\bar{p})e^{\ii\Ep t}$. Plugging this and \eqref{b-rep} into \eqref{b-evolve}, we have
\begin{align}\label{b-evolve2}
\psi_{{\rm B},a}(\vx,t)&=\sum_{s,s^\prime,b}\int Dqe^{\ii\tilde{q}\cdot\tilde{x}}\int_0^\infty\frac{dp^1}{2\pi}\frac{\mathrm{sgn}(q^1)}{2E_{p^1,\tilde{q}}}\bigl[\mathbb{M}_{ab}^s(p^1,\tilde{q},t)\mathbb{U}_1(p^1,q^1)A^{s^\prime}_qu^{s^\prime}(q)e^{\ii p^1x^1}+ \{p^1 \leftrightarrow -p^1\}\bigr] \n
	&\quad+\sum_{s,s^\prime,b}\int Dqe^{-\ii\tilde{q}\cdot\tilde{x}}\int_0^\infty\frac{dp^1}{2\pi}\frac{\mathrm{sgn}(q^1)}{2E_{p^1,-\tilde{q}}}\bigl[\mathbb{M}_{ab}^s(p^1,-\tilde{q},t)\mathbb{U}_2(p^1,q^1)B^{\dagger s^\prime}_qv^{\dagger s^\prime}(q)e^{\ii p^1x^1}+ \{p^1 \leftrightarrow -p^1\}\bigr],
\end{align}
where $E_{p^1,\tilde{q}}=\sqrt{(p^1)^2+|\tilde{q}|^2+m^2}$, and $\mathbb{U}_1(p^1,q^1)$ and $\mathbb{U}_2(p^1,q^1)$ are given by
\begin{equation}
\mathbb{U}_1(p^1,q^1)=\frac{\ii}{p^1-q^1-\ii 0}+\frac{\ii}{p^1-q^1+\ii 0},\quad \mathbb{U}_2(p^1,q^1)=\frac{\ii}{p^1+q^1-\ii 0}+\frac{\ii}{p^1+q^1+\ii 0}.
\end{equation}
In order to evaluate these integrals, we need to deform the contours of $p^1$-integral to either $(0,\ii\infty)$ or $(0,-\ii\infty)$. To which direction we deform the contours is determined in such a way that there is no contribution at infinity, and depends on the sign of $p^1$ and $x^1$ in exponentials (see Appendix C in \cite{doyonKG} for more detailed expositions). For instance, when we compute the first term of the first line in \eqref{b-evolve2} we have $e^{-\ii\Ep t+\ii p^1x^1}$ and $e^{\ii\Ep t+\ii p^1x^1}$ for which we deform the contour of the $p^1$-integral to $(0,\ii\infty)$ and $(0,-\ii\infty)$, respectively. Hence we need to evaluate only a single pole at $p^1=q^1$ which gives rise to the terms that contain $\sum_bu^s_a(q^1,\tilde{q})u^{s\dagger}_b(q^1,\tilde{q})u^{s^\prime}_b(q^1,\tilde{q})=2\Eq u^s_a(q^1,\tilde{q})\delta^{ss^\prime}$ and $\sum_bv^s_a(-q^1,-\tilde{q})v^{s\dagger}_b(-q^1,-\tilde{q})u^{s^\prime}_b(q^1,\tilde{q})=0$, thanks to the fact that $u$'s and $v$'s are orthogonal: $\sum_bv^{s\dagger}_b(-\vp)u^{s^\prime}_b(\vp)=0$. Likewise, we can extract contributions from poles $p^1=\pm q^1$ in other terms, and we end up with the following
\begin{equation}
\psi_{{\rm B},a}(\vx,t)=\int Dp\sum_s
 (\Aps u^s_a(p)e^{-\ii p\cdot x}+\Bdps v^s_a(p)e^{\ii p\cdot x}) + {\rm integral\ contribution}.
\end{equation}
In the exactly same manner as in \cite{doyonKG}, we can show that this integral contribution provides no contribution when we take the average of any observable $\mathcal{O}$ that involves $\psi(\vx,t), \psi^\dagger(\vx,t)$, and their derivatives, at $t\to \infty$.

\section{Large-$t$ behavior of oscillatory terms}\label{osci}
One can explicitly show that the oscillatory that appeared in computing matrix elements terms do not
contribute to results, i.e. decay under $t\to \infty$. Following \cite{doyonKG}, We exemplify it by calculating one of
them which appeared in (\ref{eq:osc}):
\begin{equation}
 \int_0^{\me}\frac{d\rho}{\sqrt{\me^2-\rho^2}}\frac{e^{\ii t(\Epp-\sqrt{\me^2-\rho^2})}}{\sqrt{\me^2-\rho^2}-\Epp}=\int_0^{\me}\frac{dr}{\sqrt{\me^2-r^2}}\frac{e^{\ii t(r-\Epp)}}{r-\Epp}.
\end{equation}
Other terms might be treated in a same fashion. We shall use
contour deformations again to make the asymptotic analysis feasible. First
we start with a rectangular on which the integral is performed in a complex plane parametrized by
$z$. Vertical lines of the rectangular are located at, say
$r=\me+\ii\mathrm{Im}(z)$ and $\ii\mathrm{Im}(z)$ with $0\leq \mathrm{Im}(z) \leq v$, whereas horizontal
lines are $[0,\me]$ and $\mathrm{Re}(z)+\ii v$ with $\mathrm{Re}(z)\in
[0,\me]$. Taking $v\to \infty$, the integral along the upper
horizontal line vanishes. Thus changing variables properly, we have
\begin{equation}
 \int_0^{\me}\frac{dr}{\sqrt{\me^2-r^2}}\frac{e^{\ii tr}}{r-\Epp}=-\ii\int_0^{\infty}\frac{due^{-ut}}{\sqrt{u^2-2\ii\me
   u}}\frac{e^{\ii\me
  t}}{\me+\ii u-\Epp}+\ii\int_0^{\infty}\frac{du}{\sqrt{\me^2+u^2}}\frac{e^{-ut}}{\ii u-\Epp}.
\end{equation}
For a large $t$, main contributions can be ontained by expanding
integrands around $u=0$:
\begin{align}
  \int_0^{\infty}\frac{due^{-ut}}{\sqrt{u^2-2\ii\me u}}\frac{e^{\ii\me t}}{\me+\ii u-\Epp}
  &\approx
 \frac{e^{\ii\me t}}{(\me-\Epp)\sqrt{2\ii\me }}\int_0^{\infty}du\frac{e^{-ut}}{u}\Bigl(1+\frac{u}{4\ii\me}\Bigr)\Bigl(1-\frac{\ii u}{\me-\Epp}\Bigr)
 \nonumber \\
 &=\frac{e^{\ii\me t}}{(\me-\Epp)\sqrt{2\ii\me }}t^{-\frac{1}{2}}+\mathcal{O}(t^{-1}),
\end{align}
\begin{align}
 \int_0^{\infty}\frac{du}{\sqrt{\me^2+u^2}}\frac{e^{-ut}}{\ii u-\Epp}&\approx
 -\frac{1}{\me\Epp}\int_0^{\infty}due^{-ut}\Bigl(1-\frac{u^2}{2\me^2}\Bigr)\Bigl(1+\frac{\ii u}{\Epp}\Bigr)
 \nonumber \\
 &=-\frac{1}{\me\Epp}t^{-1}+\mathcal{O}(t^{-2}).
\end{align}
Combining everything together, under $t\to \infty$, we find that the
oscillatory term decays algebraically with tails
\begin{equation}
 \int_0^{\me}\frac{d\rho}{\sqrt{\me^2-\rho^2}}\frac{e^{\ii t(\Epp-\sqrt{\me^2-\rho^2})}}{\sqrt{\me^2-\rho^2}-\Epp}=-\ii\frac{e^{\ii\me
  t}}{(\me-\Epp)\sqrt{2\ii\me}}t^{-\frac{1}{2}}+\mathcal{O}(t^{-1}).
\end{equation}
\section{Asymptotics}\label{asymp}
In this appendix we evaluate a logarithm of the determinant
$\log\mathrm{det}_{\mathcal{H}_{1\mathrm{P}}}(1+AB(t))$ where one-particle operators $A$ and $B(t)$ have matrix elements
\begin{equation}
  \langle \vp|A|\vq\rangle =A(\vp)2\Ep\,\deltasss,\ \ \langle \vp|B(t)|\vq\rangle =B(\vp,\vq)2\Ep\,\deltas.
\end{equation}
Remember that a logarithm of the determinant can be expressed as
\begin{align}\label{logdet}
 \log \mathrm{det}_{\mathcal{H}_{1\mathrm{P}}}(1+AB(t)) &= \sum_{k=1}^\infty\frac{(-1)^{k-1}}{k}\mathrm{Tr}_{\mathcal{H}_{1\mathrm{P}}}\bigl[(AB(t))^k\bigr] \n
 &=\mathrm{tr}\int\frac{d^3p}{(2\pi)^32\Ep}\langle \vec{p}|(AB(t))^k|\vec{p}\rangle.
\end{align}
where a trace $\mathrm{tr}$ is over the internal space, i.e. spins and charges. Therefore we need to evaluate
\begin{align}\label{tr}
  \int\frac{d^3p}{(2\pi)^32\Ep}\langle \vp|(AB(t))^k|\vp\rangle &=
  (2\pi)^2\delta^{(2)}(0)\int\frac{d^2p}{(2\pi)^2}\int dp_1^1\cdots
  \int dp_k^1 \nonumber \\
 &\quad \times \prod_{j=1}^k\mathbb{A}_{\tilde{p}}(p_j^1)\mathbb{B}_{\tilde{p}}(p_j^1,p_{j+1}^1)\delta_t(E_{\vp_j}-E_{\vp_{j+1}}),
\end{align}
where $\mathbb{A}_{\tilde{p}}(p_j^1):=\left.A(\vp_j)\right|_{\tilde{p}_j=\tilde{p}}$ and $\mathbb{B}_{\tilde{p}}(p_j^1,p_{j+1}^1):=\left.B(\vp_j,\vp_{j+1})\right|_{\tilde{p}_j=\tilde{p}=\tilde{p}_{j+1}}$.
Notice that the awkward term $(2\pi)^2\delta^{(2)}(0)$ is interpreted as a (infinite)
transverse area $L^2$.

When evaluating the RHS of \eqref{tr}, it is convenient to work with a variable $p_j^0=E_{\vec{p}_j}$ rather than $p^1$. However, as a map $p^1\mapsto p^0$ is not a bijection, we need to decompose the integral region of each $p_j^1$'s integral into $[-\infty,0]$ and $[0,\infty]$, which gives rise to $2^k$ $k$-tuple integrals $\int dp^1_1\cdots\int dp^1_k$ where the integral domain of each integral is either $[-\infty,0]$ or $[0,\infty]$. Of course integrals over $[-\infty,0]$ can always be transformed to that over $[0,\infty]$ by $p^1_j\mapsto-p^1_j$. Upon doing so, integrands of resulting $k$-tuple integrals can take two possible forms: one is those which consist of only $\mathbb{B}$'s whose two entries have a {\it same} sign, i.e. either $\mathbb{B}_{\tilde{p}}(p_j^1,p_{j+1}^1)$ or $\mathbb{B}_{\tilde{p}}(-p_j^1,-p_{j+1}^1)$. Notice that there are two such $k$-tuple integrals. Another case is those which contain at least one $B$'s whose two entries have {\it opposite} signs like $\mathbb{B}_{\tilde{p}}(-p_j^1,p_{j+1}^1)$. This is the dominant case that constitutes $2^k-2$ $k$-tuple integrals out of $2^k$ ($k$-tuple) integrals. Having $p^1_j$ integrals over $[0,\infty]$, we can change the integration variable to $p^0_j$, and expand $\mathbb{
B}_{\tilde{p}}(\pm p_j^1,\pm p_{j+1}^1)$ around
$p_j^0=p_{j+1}^0$, yielding
\begin{equation}\label{bexp}
 \mathbb{B}_{\tilde{p}}(\pm p_j^1,\pm p_{j+1}^1)= \mathbb{B}_{\tilde{p}}(\pm p^1_j, \pm p^1_j)+\sum_{n>0}c_n(p_j^0,\tilde{p})(p_j^0-p_{j+1}^0)^n,
\end{equation}
with coefficients $c_n(p^0_j,\tilde{p})$. As we shall see below, however, terms that contain higher powers of $p^0_j-p^0_{j+1}$ (second term in \eqref{bexp}) do not contribute to the leading order: their contribution is of order $\mathcal{O}(1)$ in time $t$, and suppressed by the ballistic contributions (linear in $t$) by $\mathbb{B}_{\tilde{p}}(\pm p^1_j, \pm p^1_j)$. Furthermore, recalling that, in our application, $\mathbb{B}_{\tilde{p}}(p^1,q^1)\propto \bar{u}(p^1,\tilde{p})\gamma^\mu u(q^1,\tilde{p})$ (or $\bar{v}(p^1,\tilde{p})\gamma^\mu v(q^1,\tilde{p})$), it follows from  the Gordon identity
\begin{equation}
\bar{u}(p)\gamma^\mu u(q)=\bar{u}(p)\Bigl(\frac{p^\mu+q^\mu}{2}-\frac{[\gamma^\mu,\gamma^\nu](p-q)_\nu}{4m}\Bigr)u(q)
\end{equation}
that $\mathbb{B}_{\tilde{p}}(p^1_j, -p^1_j)=0=\mathbb{B}_{\tilde{p}}(-p^1_j, p^1_j)$. Therefore in such a  situation, it turns out that only 2 out of $2^k$ $k$-tuple integrals, which are made of either only $\mathbb{B}_{\tilde{p}}(p_j^1,p_{j+1}^1)$ or $\mathbb{B}_{\tilde{p}}(-p_j^1,-p_{j+1}^1)$, have non-vanishing contributions to the leading order. Let us focus on this special case hereafter as the application we have in mind belongs to this situation which makes arguments substantially simplified. We further deal with a $k$-tuple integral in which only $\mathbb{B}_{\tilde{p}}(p_j^1,p_{j+1}^1)$ appear: another case is completely analogous to this one. Defining $\mathbb{B}^+_{\tilde{p}}(p^0_j):=\mathbb{B}_{\tilde{p}}(p^1_j,p^1_j)$ and $\mathbb{A}^+_{\tilde{p}}(p^0_j):=\mathbb{A}_{\tilde{p}}(p^1_j)$, the main part
of \eqref{tr} can be divided into two parts
\begin{equation}
 \int_0^{\infty}dp_1^1\cdots
  \int_0^{\infty}dp_k^1\prod_{j=1}^k\mathbb{A}_{\tilde{p}}(p_j^1,\tilde{p})\mathbb{B}(p_j^1,p_{j+1}^1,\tilde{p})\delta_t(E_{\vp_j}-E_{\vp_{j+1}})=V^+(\me)+W^+(\me),
\end{equation}
where
\begin{align}
V^+(\me)&=\int_{\me}^{\infty}dp_1^0\cdots
  \int_{\me}^{\infty}dp_k^0\prod_{j=1}^k\mathbb{A}^+_{\tilde{p}}(p_j^0)\mathbb{B}^+_{\tilde{p}}(p_j^0)\frac{p_j^0}{p_j^1}\delta_t(p_j^0-p_{j+1}^0) \\
 W^+(\me)&=\sum_{l,m>0}\int_{\me}^{\infty}dp_1^0\cdots\int_{\me}^{\infty}dp_k^0\prod_{j=1}^k\mathbb{A}^+_{\tilde{p}}(p_j^0)\frac{p_j^0}{p_j^1}C_{lm}(p_1^0,\cdots,p_k^0,\tilde{p})(p_l^0-p_{l+1}^0)^m\delta_t(p_j^0-p_{j+1}^0), \label{eq:sum}
\end{align}
with $\me=\me(\tilde{p})$ satisfying $p^0_j=\sqrt{(p^1_j)^2+\me^2}$ for
any $j=1,\cdots, k$.
We emphasize that each term in the summation in (\ref{eq:sum}) can always be expressed as a term like
$(p_l^0-p_{l+1}^0)^m$ with a coefficient depending possibly on all
energies $p_1^0,\cdots,p_k^0$ - the summation can be written in any way as long
as this is indicated. Let us recall that the following relation shown
in \cite{BD1}: for any domain $\mathcal{S} \subset \mathbb{R}$ and
endomorphism $f$ of $\mathbb{R}$
\begin{equation}\label{deltat}
 \lim_{t\to \infty}\int_{\mathcal{S}}dp_1\cdots
  \int_{\mathcal{S}}dp_k\prod_{j=1}^kf(p_j)\delta_t(p_j-p_{j+1}) = t\int_{\mathcal{S}}\frac{dp}{2\pi}f(p)^k+\mathcal{O}(1),
\end{equation} 
or equivalently, after the Fourier transformation,
\begin{equation}
 \lim_{t\to \infty}\int_{\mathcal{S}}dp_1\cdots
  \int_{\mathcal{S}}dp_ke^{\ii\sum_{j=1}^k\alpha_jp_j}\prod_{j=1}^k\delta_t(p_j-p_{j+1}) = t\int_{\mathcal{S}}\frac{dp}{2\pi}e^{\ii\sum_{j=1}^k\alpha_jp}+\mathcal{O}(1).
\end{equation}
Applying this to our case, we can readily see that
$W(\me)=0$. Concretely, we first Fourier transform
$A^+(p_j^0)p_j^0/p_j^1$ as $A^+(p_j^0)p_j^0/p_j^1=\int
d\alpha_jA^+_{\alpha_j}e^{\ii\alpha_jp_j^0}$, then there exists a linear combination of (products of) differential
operators $\mathcal{D}_{lm}(\alpha_1,\cdots,\alpha_k)$ which acts on
$e^{\ii\alpha_jp_j^0}$ and produces a coefficient
$C_{lm}$. Furthermore the action of a differential operator
$(-\ii)^n(\partial_l-\partial_{l+1})^n$ on $e^{i\sum_{j=1}^k\alpha_jp_j^0}$ results in
$e^{\ii\alpha_jp_j^0}(p_l^0-p_{l+1}^0)^n$. Hence now we can use the
aforementioned formula:
\begin{align}
 W(\me)&=\int d\alpha_1\cdots\int d\alpha_k\sum_{l,m>0}\mathcal{D}_{lm}(-\ii)^n(\partial_l-\partial_{l+1})^n\int_{\me}^{\infty}dp_1^0\cdots\int_{\me}^{\infty}dp_k^0e^{\ii\sum_{j=1}^k\alpha_jp_j^0}\delta_t(p_j^0-p_{j+1}^0)
 \nonumber \\
&=t\int d\alpha_1\cdots\int d\alpha_k\sum_{l,m>0}\mathcal{D}_{lm}(-\ii)^n(\partial_l-\partial_{l+1})^n\int_{\me}^{\infty}\frac{dp}{2\pi}e^{\ii\sum_{j=1}^k\alpha_jp}
 \nonumber \\
&=t\int d\alpha_1\cdots\int d\alpha_k\sum_{l,m>0}\mathcal{D}_{lm}\int_{\me}^{\infty}\frac{dp}{2\pi}e^{\ii\sum_{j=1}^k\alpha_jp}(p-p)^n
 \nonumber \\
 &=0.
\end{align}
Thus we find that in fact only $V(\me)$ contributes to the
determinant. A similar argument also holds for a $k$-tuple integral that contains only $\mathbb{B}_{\tilde{p}}(-p^1_j,-p^1_{j+1})$, and correspondingly we define $\mathbb{B}^-_{\tilde{p}}(p^0_j):=\mathbb{B}_{\tilde{p}}(-p^1_j,-p^1_j), \mathbb{A}^-_{\tilde{p}}(p^0_j):=\mathbb{A}_{\tilde{p}}(-p^1_j)$, and $V^-(\me)$ in a same manner as above. We then again employ the formula \eqref{deltat} to these $V^+(\me)$ and $V^-(\me)$, obtaining
\begin{align}
\int \frac{d^3p}{(2\pi)^32\Ep}\langle \vp|(AB(t))^k|\vp\rangle &=
  t(2\pi)^2\delta^{(2)}(0)\int\frac{d^2p}{(2\pi)^2}\int_{\me}^{\infty}\frac{dp^0}{2\pi}\biggl[\mathbb{A}^+_{\tilde{p}}(p^0)^k\Bigl(\mathbb{B}^+_{\tilde{p}}(p^0)\frac{p^0}{p^1}\Bigr)^k \n
  &\quad +\mathbb{A}^-_{\tilde{p}}(p^0)^k\Bigl(B^-_{\tilde{p}}(p^0)\frac{p^0}{p^1}\Bigr)^k\biggr]+\mathcal{O}(1).
\end{align}
Taking the trace over the internal space into account, the result of the
whole trace reads
\begin{align}
\mathrm{Tr}_{\mathcal{H}_{1\mathrm{P}}}\bigl[(AB(t))^k\bigr]&=2t(2\pi)^2\delta^{(2)}(0)\sum_{c=\pm}\int\frac{d^2p}{(2\pi)^2}\int_{\me}^{\infty}\frac{dp^0}{2\pi}\biggl[\mathbb{A}_{\tilde{p},c}^+(p^0)^k\Bigl(\mathbb{B}_{\tilde{p},c}^+(p^0)\frac{p^0}{p^1}\Bigr)^k \n
  &\quad +\mathbb{A}_{\tilde{p},c}^-(p^0)^k\Bigl(\mathbb{B}_{\tilde{p},c}^-(p^0)\frac{p^0}{p^1}\Bigr)^k\biggr] +\mathcal{O}(1),
\end{align}
and we finally have the desired asymptotic behavior under $t\to \infty$
\begin{align}
 \log \mathrm{det}_{\mathcal{H}_{1\mathrm{P}}}(1+AB(t)) &=
  2t(2\pi)^2\delta^{(2)}(0)\sum_{c=\pm}\int\frac{d^2p}{(2\pi)^2}\int_{\me}^{\infty}\frac{dp^0}{2\pi}\Bigl[\log(1+\mathbb{A}^+_{\tilde{p},c}(p^0)\mathbb{B}^+_{\tilde{p},c}(p^0)\frac{p^0}{p^1})
 \nonumber \\
&\quad+\log(1+\mathbb{A}^-_{\tilde{p},c}(p^0)\mathbb{B}^-_{\tilde{p},c}(p^0,\tilde{p})\frac{p^0}{p^1})\Bigr]+\mathcal{O}(1),
\end{align}
where $\langle c^{\prime} |X|c \rangle=X_c\delta_{c^{\prime},c}$ for
$X=\mathbb{A}^\pm,\mathbb{B}^\pm$.

\section{Integration}\label{integ}
Here we compute (\ref{eq:ll}). We first note that the following integration formula readily follows after straightforward calculations: for $a, c\in \mathbb{R}$ and $b\geq0$,
\begin{equation}
 \sum_{\epsilon=\pm}\int_b^{\infty}\frac{dp\epsilon}{e^{p-\epsilon(a+\ii c)}+1}=
\begin{dcases*}								   
-2\sum_{n=1}\frac{(-1)^n}{n}e^{-bn}\sinh[(a+\ii c)n] & $|a|<b$ \\
a-b-2\sum_{n=1}^{\infty}\frac{(-1)^n}{n}\Bigl[\ii\sin(cn)+e^{-(a+\ii c)n}\sinh(bn)\Bigr]
 & $a>b$ \\
a+b-2\sum_{n=1}^{\infty}\frac{(-1)^n}{n}\Bigl[\ii\sin(cn)-e^{(a+\ii c)n}\sinh(bn)\Bigr]
 & $a<-b$
\end{dcases*}.
\end{equation}
Thus defining $\bar{f}(\lambda,\tilde{p}S;\beta,\mu)$ by
\begin{equation}
 \bar{f}(\lambda,\tilde{p};\beta,\mu)=\sum_{\epsilon=\pm}\int_{\me({\tilde{p}})}^{\infty}\frac{dE}{2\pi}\log\bigl[1+n_{\epsilon}(e^{\ii\epsilon
  \lambda}-1)\bigr],
\end{equation}
we can express $\bar{J}(\lambda,\tilde{p};\beta,\mu):=d\bar{f}(\lambda,\tilde{p};\beta,\mu)/d(\ii\lambda)$ as
\begin{align}
 \bar{J}(\lambda,\tilde{p};\beta,\mu)&=\beta^{-1}\sum_{\epsilon=\pm}\int_{\beta\me({\tilde{p}})}^{\infty}\frac{dp}{2\pi}\frac{\epsilon}{e^{p-\epsilon(\beta\mu+\ii\lambda)}+1}
 \nonumber \\
&=\begin{dcases*}
   -\frac{1}{\pi\beta}\sum_{n=1}^{\infty}\frac{(-1)^n}{n}e^{-\beta
   \me(\tilde{p})n}\sinh[(\beta \mu+\ii\lambda)n] & $|\mu|<\me({\tilde{p}})$ \\
   \frac{\mu-\me(\tilde{p})}{2\pi}-\frac{1}{\pi\beta}\sum_{n=1}^{\infty}\frac{(-1)^n}{n}\Bigl[\ii\sin(\lambda
 n)+e^{-(\beta \mu+\ii\lambda)n}\sinh(\beta \me(\tilde{p})n)\Bigl] & $\mu>\me({\tilde{p}})$ \\
 \frac{\mu+\me(\tilde{p})}{2\pi}-\frac{1}{\pi\beta}\sum_{n=1}^{\infty}\frac{(-1)^n}{n}\Bigl[\ii\sin(\lambda
 n)-e^{(\beta \mu+\ii\lambda)n}\sinh(\beta \me(\tilde{p})n)\Bigl] & $\mu<-\me({\tilde{p}})$
  \end{dcases*}.
\end{align}
Upon performing integrals first over the transverse momenta $\lambda$, we
have (\ref{eq:ll2}) and (\ref{eq:ll3}).

\section{Energy fluctuations}\label{energy}
It would be natural to attempt the generalization of the above result to the
energy transport as well. Here we assume that the EFR for the energy
transport
\begin{equation}
 G(\sigma)=\int_0^{\ii\sigma}dy J_E(\betal-y,\betal \mul;\betar+y,\betar \mur)
\end{equation}
holds as does \cite{doyonKG}. We can then readily show
that an average of the energy current $J_E(\betal,\betal
\mul;\betar,\betar \mur)$ in the NESS is
\begin{align}
J_E(\betal,\betal
\mul;\betar,\betar \mur)= \langle T^{01}\rangle_{\mathrm{s}} &
= 2\int \Dp p^1\Bigl(\frac{1}{e^{V_+({\bf p})}+1}+ \frac{1}{e^{V_-({\bf
 p})}+1}\Bigr) \nonumber \\
&=
 \frac{3}{2\pi^2}\sum_{\epsilon=\pm}\biggl(\zeta_{\frac{m}{T_{\mathrm{L}}},\frac{\epsilon
 \mu_{\mathrm{L}}}{T_{\mathrm{L}}}}(4)T_{\mathrm{L}}^4-\zeta_{\frac{m}{T_{\mathrm{R}}},\frac{\epsilon
 \mu_{\mathrm{R}}}{T_{\mathrm{R}}}}(4)T_{\mathrm{R}}^4 \biggr),
\end{align}
where the stress-energy tensor for the Dirac model is given by
\begin{equation}
 T^{\mu \nu} = :\ii\overline{\psi}\gamma^{\mu}\partial^{\nu}\psi-\eta^{\mu
  \nu}[\overline{\psi}(\ii\partial_{\sigma}\gamma^{\sigma}-m)\psi]:,
\end{equation}
and we defined
\begin{equation}
 \zeta_{a,b}(4)=\frac{1}{\Gamma(4)} \int_0
  ^{\infty}dp\frac{p^3}{e^{\sqrt{p^2+a^2}-b}+1}.
\end{equation}
From now on we shall discuss only the massless case for simplicity: the
extension to the massive case is straightforward. In the massless limit
$m=0$, the result becomes rather concise. The average energy current and the
associated SCGF are simply
\begin{align}
 J_E(\betal,\betal
\mul;\betar,\betar \mur)
 &=\frac{\mu^4_{\mathrm{L}}-\mu^4_{\mathrm{R}}+2\pi^2(\mu^2_{\mathrm{L}}T^2_{\mathrm{L}}-\mu^2_{\mathrm{R}}T^2_{\mathrm{R}})+\frac{7\pi^2}{15}(T^4_{\mathrm{L}}-T^4_{\mathrm{R}})}{16\pi^2} \\
 G(\sigma)&=g(\sigma,\betal,\betal\mul)+g(-\sigma,\betar,\betar\mur)
\end{align}
with
\begin{equation}
 g(\sigma,\beta,\beta \mu)=\biggl(\frac{7}{45}\pi^2+\frac{(\mu\beta)^2}{24}+\frac{(\mu\beta)^4}{48\pi^2}\biggr)\biggl(\frac{1}{(\beta-\ii\sigma)^3}-\frac{1}{\beta^3}\biggr).
\end{equation}
If we set $\mu=0$, then this SCGF is similar to that for the
Klein-Gordon theory \cite{doyonKG} up to its coefficient, and hence
can be interpreted via Poisson processes. Unlike the charge transport,
this is valid for the entire domain $\sigma \in \mathbb{R}$ since the
charge quantization does not affect the energy transfer in a sense that
the SCGF has no $2\pi$ periodicity. In a same fashion as
\cite{BD2} it is also possible to derive the SCGF
$H(\lambda,\sigma)$ for both transports that satisfies the following set
of PDEs
\begin{align}
 \frac{\partial H(\lambda,\sigma)}{\partial (\ii\sigma)} &= J_E(\betal-\ii\sigma,
 \betal\mul+\ii\lambda;\betar+\ii\sigma, \betar\mur-\ii\lambda) \label{eq:pde1}\\
\frac{\partial H(\lambda, \sigma)}{\partial (\ii\lambda)} &= J_Q(\betal-\ii\sigma,
 \betal\mul+\ii\lambda;\betar+\ii\sigma, \betar\mur-\ii\lambda). \label{eq:pde2}
\end{align}
It is a simple matter to confirm that a consistency condition
\begin{equation}
 \frac{\partial^2H(\lambda,\sigma)}{\partial (\ii\sigma) \partial (\ii\lambda)} =
  \frac{\partial^2H(\lambda,\sigma)}{\partial (\ii\lambda) \partial (\ii\sigma)}
\end{equation}
is met. The total SCGF which combines both charge and energy
transfers is, for $\lambda \in (-\pi,\pi)$,
$H(\lambda,\sigma)=h(\lambda,\sigma;\betal,\betal\mul)+h(-\lambda,-\sigma;\betar,\betar\mur)$
where
\begin{equation}
 h(\lambda,\sigma;\beta,\beta\mu)=\frac{1}{48\pi^2}\biggl[\frac{(\beta\mu+\ii\lambda)^4+2\pi^2(\beta
  \mu+\ii\lambda)^2+\frac{7}{30}\pi^4}{(\beta-\ii\sigma)^3}-\frac{(\beta \mu)^4+2\pi^2(\beta\mu)^2+\frac{7}{30}\pi^4}{\beta^3}\biggr].
\end{equation}

\section{Lifshitz fermions}\label{lifshitz}
Non-equilibrium charge transports in non-relativistic free systems might
be also treated in the exactly same fashion as in the Dirac theory. Here we generalize our approach to the Lifshitz-type free fermion model whose Hamiltonian
reads
\begin{equation}
 H=\int \frac{d^3p}{(2\pi)^3}\sum_s\Op c^{\dagger s}_{\vp}c^s_{\vp}
\end{equation}
where $\Op=|\vp|^{z}/2m$ for $z=1,2,\cdots$. $c^{\dagger s}_{\vp}$ and $c^s_{\vp}$ satisfy the previous anticommutation relation
\eqref{commu}. For
$z=2$, this model is nothing but a non-relativistic free fermion
system. Assuming the NESS density matrix for this model
has a similar form as \eqref{ness}, the average current in the NESS $\mathcal{J}(\betal,\betal\mul;\betar,\betar\mur)$ is then
given by
\begin{align}
 \mathcal{J}(\betal,\betal\mul;\betar,\betar\mur)&=2\int_{p^1>0}\frac{d^3p}{(2\pi)^3}
 \frac{d\Op}{dp^1}\biggl(\frac{1}{e^{\betal(\Op-\mul)}+1}-\frac{1}{e^{\betar(\Op-\mur)}+1}\biggr)\nonumber \\
 &=\frac{z}{8\pi^2m}\int_0^{\infty}dp\biggl(\frac{p^{z+1}}{e^{\betal(\Op-\mul)}+1}
 -\{\mathrm{L} \leftrightarrow \mathrm{R} \}\biggr) \nonumber \\
&:=\mathcal{I}(\betal,\betal\mul)-\mathcal{I}(\betar,\betar\mur),
\end{align}
where
\begin{equation}
 \mathcal{I}(\beta,\beta\mu)=\frac{1}{8\pi^2m}\Bigl(\frac{2m}{\beta}\Bigr)^{1+\frac{2}{z}}\int_0^{\infty}dk\frac{k^{\frac{2}{z}}}{e^{k-\beta\mu}+1}.
\end{equation}
One might notice that, in terms of the polylogarithm, this can be
expressed as
\begin{equation}
 \mathcal{I}(\beta,\beta\mu)=-\frac{1}{8\pi^2m}\Bigl(\frac{2m}{\beta}\Bigr)^{1+\frac{2}{z}}\Gamma\Bigl(1+\frac{2}{z}\Bigr)\mathrm{Li}_{1+\frac{2}{z}}\bigl(-e^{\beta\mu}\bigr).
\end{equation}
In the low temperature regime ($\beta \gg 1$), this has an asymptotic
expansion
\begin{equation}
 \mathcal{I}(\beta,\beta\mu)=\frac{(2m)^{\frac{2}{z}}}{4\pi^2}\biggl[\frac{z}{2+z}\mu^{1+\frac{2}{z}}+\frac{\pi^2}{3z}\beta^{-2}\mu^{-1+\frac{2}{z}}+\mathcal{O}(\beta^{-4})\biggr].
\end{equation}
If we set $z=2$, this recovers the result obtained in
\cite{CM}. Furthermore by means of the EFR we can compute the SCGF
for this charge transport
$\mathcal{F}(\lambda)=\mathcal{G}(\lambda;\betal,\betal\mul)+\mathcal{G}(-\lambda;\betar,\betar\mur)$
immediately as 
\begin{equation}
 \mathcal{G}(\lambda,\beta,\beta\mu)=-\frac{1}{8\pi^2m}\Bigl(\frac{2m}{\beta}\Bigr)^{1+\frac{2}{z}}\Gamma\Bigl(1+\frac{2}{z}\Bigr)\biggl[\mathrm{Li}_{2+\frac{2}{z}}\bigl(-e^{\beta\mu+i\lambda}\bigr)-\mathrm{Li}_{2+\frac{2}{z}}\bigl(-e^{\beta\mu}\bigr)\biggr].
\end{equation}
The extension of the above computation to generic dimensions is
straightforward.


\end{document}